\documentclass[aps,pra,twocolumn,a4paper,amsmath,amsthm,amssymb,showpacs,%
floatfix,superscriptaddress]{revtex4}
\usepackage{braket}
\usepackage{graphicx}
\usepackage{dcolumn}
\usepackage{dsfont}
\usepackage{epstopdf}
\usepackage{xcolor}
\usepackage[normalem]{ulem}

\begin{document}

\title{Precisely controlling the reflection phase of a photon via a strongly-coupled ancilla dressed qubit}

\author{F. Motzoi}
\author{K. M\o lmer}
\affiliation{Department of Physics and Astronomy, Aarhus University, Aarhus, Denmark}
\date{\today}

\begin{abstract}
We propose that Rydberg dressing of a single qubit atom can be used to control a surrounding ensemble of three-level atoms and hereby the phase of light reflected by an optical cavity. Our scheme employs an ensemble dark resonance that is perturbed by the qubit state of a single atom to yield a single-atom single-photon gate. We show here that off-resonant Rydberg dressing of the qubit offers experimentally-viable regimes of operation that drastically reduce error compared to schemes using shelved Rydberg population. Such low errors (in the $10^{-3}$ range) are a necessary condition for fault-tolerant optical-photon, gate-based quantum computation.  We also demonstrate the technique for microwave circuit-QED, where a strongly-coupled ancilla superconducting qubit can be used in the place of the atomic ensemble to provide high-fidelity coupling to microwave photons.
\end{abstract}

\maketitle

\section{Introduction}

The transmission of quantum information between remote quantum systems represents one of the main technical bottlenecks to the scalability of quantum networks for distributed quantum computing, cryptography, metrology and sensing \cite{kimble_quantum_2008}. Proposals to use light to interlink quantum degrees of freedom of spatially separated nodes fall broadly in two categories. The first engages direct transmission of non-classical states of light \cite{duan_scalable_2004,kimble_quantum_2008,reiserer_nondestructive_2013,reiserer_quantum_2014,hao_quantum_2015,motzoi_backaction-driven_2016,das_photonic_2016}, while the second heralds non-local quantum correlations by joint measurements on signals that are emitted from or have sequentially interacted with spatially separated quantum systems \cite{firstenberg_nonlinear_2016,cabrillo_creation_1999-1,moehring_entanglement_2007,hensen_loophole-free_2015,roch_observation_2014,motzoi_continuous_2015,martin_deterministic_2015,duan_quantum_2000,duan_long-distance_2001}.

In all cases it is pertinent to have an efficient coupling of the matter and light degrees of freedom, which can be achieved when high Q cavities are used to enhance the coupling of even a single atom with quantum light \cite{pellizzari_quantum_1997,wilk_single-atom_2007}, and when photons interact with the collective quantum degrees of freedom of large ensembles of atoms. In the former case, the matter-light interaction has enabled quantum gates on the atomic and photonic qubits \cite{reiserer_quantum_2014}, while in the latter, long-range interactions between the atoms have been used to establish effective optical non-linearities, e.g., when a delocalized single Rydberg excitation disrupts the propagation of slow light by electromagnetically induced transparency (EIT) through the surrounding medium \cite{ates_electromagnetically_2011,firstenberg_nonlinear_2016,chang_quantum_2014,tiecke_nanophotonic_2014}.

Ensembles of interacting atoms can be employed to create and manipulate non-classical states of light \cite{lukin_dipole_2001,saffman_creating_2002,ningyuan_observation_2016,ding_entanglement_2016} to enable qubit interactions between separate single photon wave packets \cite{das_photonic_2016,hao_quantum_2015}. It is also possible to address collective qubit degrees of freedom via different internal atomic states \cite{pedersen_few_2009,grankin_quantum-optical_2015}, but effective coupling of single photons to single atomic qubits that form part of a local register with computing, memory or sensing capabilities remains a challenge. Thus far, experimental singe-atom cavity QED \cite{reiserer_quantum_2014} has achieved errors in the 10\% range with high-Q cavities, while proposals based on strongly-coupled Rydberg ensembles suggest this can be brought down to the few \% level \cite{wade_single-atom_2016}. However, these proposals involve shelving population in the Rydberg levels which are subject to shorter lifetimes than the hyperfine ground states and to leakage to nearby Rydberg levels. This has caused Rydberg excitation gates to shelved states to be limited to the $80\%$ fidelity range \cite{maller_rydberg-blockade_2015,jau_entangling_2016}.

In this work, we build on earlier proposals that use Rydberg blockade due to shelved Rydberg states to alter the EIT condition for a large ensemble of atoms interacting with a cavity field, and thus alter the reflection property of the cavity surrounding the atoms \cite{hao_quantum_2015,das_photonic_2016,wade_single-atom_2016}; in particular we improve the single-qubit photon switch \cite{wade_single-atom_2016} where a single atomic qubit is used to block the Rydberg state of the ensemble of a different atomic species. While retaining the advantages of the collectively enhanced matter-light interaction, we explore here a different control mechanism and we show that qubit-photon gates can be achieved with little excitation of the Rydberg levels.  We also generalize the ideas presented to other physical systems such as superconducting circuit-QED with microwave optics, showing that these techniques can be applied to a variety of different matter systems that couple to single photons.

The article is organized as follows, In Sec.~II, we present the physical system, we review the cavity EIT mechanism introduced in \cite{wade_single-atom_2016}, and we present the new physical mechanism at play in our Rydberg dressing proposal. In Sec.~III, we present a complete input-output quantum analysis, including solution of the Schr\"odinger equation for the combined discrete atomic and cavity degrees  of freedom and continuous incident and reflected field modes. In Sec.~IV, we present results showing the scaling of the fidelity of a single-atom single-photon phase gate with different physical parameters. Sec. V discusses an alternative physical system for the proposal, namely in circuit-QED. Sec. VI summarizes the results of the article.

\section{The physical system and the atom-photon gate mechanism}

The physical system and atomic level schemes are depicted in Fig.1. The (red) qubit atom is selectively excited towards a Rydberg state from the $|1_q\rangle$ (hyperfine) ground state by a classical laser field with Rabi frequency $\epsilon$ and detuning $\Delta$ . For the ensemble of $N$ (blue) ancilla atoms, $G$ denotes the $\sqrt{N}$ times enhanced collective absorption amplitude of a single cavity photon, exciting the ground product state $\ket{0_a}=|0,0,..., 0\rangle_a$ to the collectively excited state, $\ket{1_a}=(|1,0,..., 0\rangle_a + |0,1,0,..., 0\rangle_a + |0,.... 1\rangle_a)/\sqrt{N}$, ($N$ terms), which, in turn, is subject to strong driving with the Rabi frequency $\Omega$ to the collectively excited Rydberg state  $\ket{2_a}=(|2,0,..., 0\rangle_a + |0,2,0,..., 0\rangle_a + |0,..., 2\rangle_a)/\sqrt{N}$.

The incident photon is resonant with the empty cavity mode and the lower optical transition in the ancilla atoms, and it thus couples resonantly to the dark state, $\ket{\psi_0} = (G \ket{0}\ket{ 2_a} - \Omega \ket{1}\ket{ 0_a})/\sqrt{G^2+\Omega^2}$, with either a single cavity photon or a shared excitation in the atoms. When the qubit atom is excited to its Rydberg state and effectively blocks excitation of the ancilla Ryderg states, a large number of ancilla atoms behave as two level-atoms and their strong coupling to the cavity mode splits the cavity resonance and prevents excitation by the incident photon.
It was shown in \cite{wade_single-atom_2016}, that due to the transfer into the dark state and back into the propagating field, the phase of the reflected photon differs by $\pi$ from the case where the photon is excluded from entering the cavity.
In free space, the disruption of EIT leads to absorption, and e.g., to anti-bunched transmission \cite{firstenberg_attractive_2013,tiecke_nanophotonic_2014,chang_quantum_2014}, while, by splitting the cavity frequency, it yields a dispersive effect for the photon scattering process (see \cite{gorshkov_dissipative_2013} for recent dispersive variants of free space Rydberg EIT).

Note that either the photon  never enters the cavity, or it enters into a mostly photonic state (if $\Omega \gg G$), where the ancilla atoms are only little excited. The qubit atom may have no optical transition in the frequency range of the photons, offering possibilities to separately optimize the qubit lifetime and transmission wavelengths of the quantum network.

A complete input-out theory was employed in \cite{wade_single-atom_2016} to assess the fidelity of the phase gate due to decay and losses and the finite bandwidth of the cavity and the dark state mechanism. A simple estimate of the loss of fidelity due to atomic decay showed that the excitation of the qubit Rydberg state during the reflection process  may, indeed, be the dominant error. This is the motivation for the present study including a Rydberg dressed, i.e., off-resonantly excited qubit atom, with a correspondingly reduced decay probability. While the phase gate mechanism is similar with the one in \cite{wade_single-atom_2016}, we discover new mechanisms and possibilities to suppress the gate error.

\begin{figure}
	\includegraphics[width=.55\textwidth]{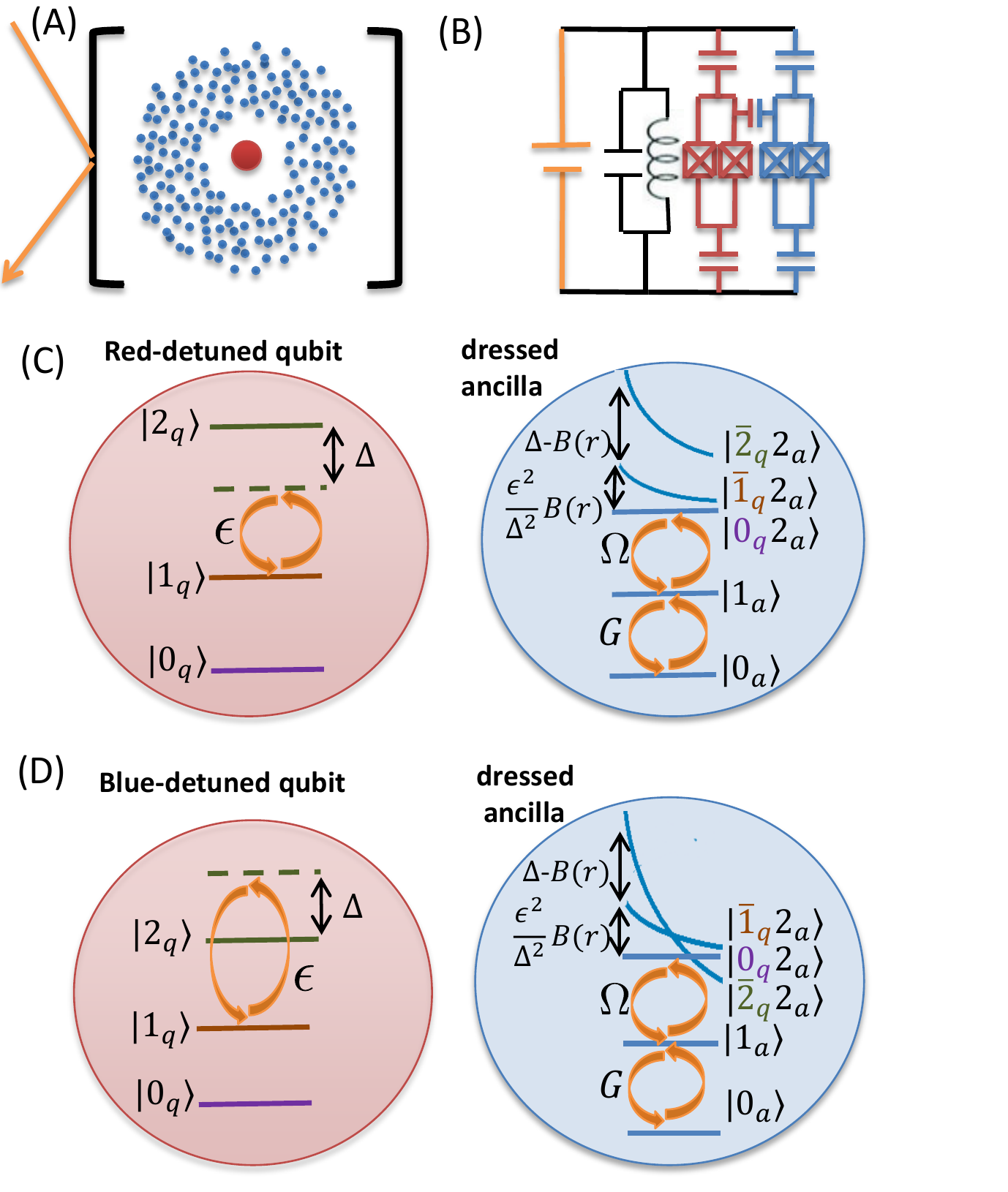}
	\caption{ Different architectures for implementing high-fidelity atom-photon gate: (A) Geometry where ancilla atoms are distributed around a central qubit atom, and (B) an equivalent circuit with superconducting artificial atoms.   The level diagram is plotted for the atom cloud distribution as a function of distance between the qubit and ancilla atoms in (C) and (D) for red and blue detuned dressing, respectively. For an excited qubit, the EIT mechanism is no longer resonant. In the blue-detuned case, the dressed Rydberg levels become resonant at some distance from the center, causing an (EIT dark state) splitting of $2\epsilon$, enhancing the photon blockade at this distance.
 \label{fig:1}}
\end{figure}

\section{Input-output theory in the Schr\"odinger picture}

As the photon may not only be reflected but its wave form may be entangled with the atoms, we need to account for its continuum wavefunction along with the amplitudes on the various discrete collective states of the cavity field and the atoms. To this end it is useful to apply input-output theory in the Sch\"odinger picture, rather than in the usually employed Heisenberg picture \cite{gardiner_driving_1993}. In the same way as the input-output theory simplifies when the coupling terms and Hamiltonians are second order in oscillator quadrature operators and the coupled Heisenberg picture equations of motion are linear, the Schr\"odinger picture equations simplify considerably, when only a single quantum of excitation is introduced and shared between the different components of  the system.

\subsection{Empty cavity}

In this section, we review the equations of motion for the amplitude on the state with an empty cavity illuminated by a one photon continuum wave packet.
We expand the state of the system containing a single photon as
\begin{equation} \label{eq:singlephoton}
|\Psi(t)\rangle = C(t)|1,0\rangle + \int d\omega \phi(\omega,t)|0,1_\omega\rangle,
\end{equation}
with  $|1,0\rangle = b^\dagger|0,0\rangle$ and $|0,1_\omega\rangle=a^\dagger(\omega)|0,0\rangle$ denoting one photon states in the cavity mode and in the field eigenmode with frequency $\omega$, respectively (we do not consider polarization degrees of freedom in this work).

We apply a rotating frame with respect to the cavity mode, so that $\omega$ denotes the field detuning with respect to the cavity, and the fields are described by the free field Hamiltonian
\begin{equation}
H_F = \hbar \int d\omega\ \omega a^\dagger(\omega)a(\omega),
\end{equation}
and by the coupling due to the mirror
\begin{equation}
H_M= i\hbar\int d\omega\ g(\omega)[a^\dagger(\omega) b - b^\dagger a(\omega)].
\end{equation}

The Schr\"odinger equation yields for $\phi(\omega,t)$,
\begin{equation} \label{eq:field}
\dot{\phi}(\omega,t) = -i\omega \phi(\omega,t) + g(\omega) C(t),
\end{equation}
which can be integrated directly, assuming the expansion of the incident photon wave packet on the field eigenmodes at time $t=0$ prior to the reflection process,
\begin{equation}
\phi(\omega,t) = e^{-i\omega t} \phi(\omega,0) + g(\omega) \int_0^t ds e^{-i\omega(t-s)} C(s).
\end{equation}

Inserting this expression in the equation for $C(t)$ and employing the Born-Markov approximation yields
 \begin{equation} \label{eq:cav}
\dot{C}(t) = - \frac{\kappa}{2} C(t) - \sqrt{\kappa} \beta_{in}(t),
\end{equation}
where we have introduced $\kappa=2\pi|g(\omega)|^2$ evaluated at the cavity eigenfrequency, and $\beta_{in}(t) = \frac{1}{\sqrt{2\pi}}\int d\omega e^{-i\omega t}\phi(\omega,0)$, which represents the time dependent arrival of the incident photon wave packet on the input mirror. Note that we only need to solve a single equation for the intracavity photon amplitude subject to a driving term $\beta_{in}(t)$, given by the shape of the incident wave packet.

Integrating (\ref{eq:field}) backwards in time from time $t=T$, long after the reflection process, yields
  \begin{equation}
\phi(\omega,t) = e^{-i\omega (t-T} \phi(\omega,T) -g(\omega)\int_t^T ds e^{-i(\omega(t-s)} C(s).
\end{equation}
Inserting this expression in the equation for $C(t)$ yields
 \begin{equation}
\dot{C}(t) = \frac{\kappa}{2} C(t) - \sqrt{\kappa} \beta_{out}(t),
\end{equation}
where $\beta_{out}(t) = \frac{1}{\sqrt{2\pi}}\int d\omega e^{-i\omega (t-T}\phi(\omega,T)$ is the time dependent shape of the photon wavepacket as it propagates away from the cavity mirror.

Subtracting the two equations for $\dot{C}(t)$ yields the input-output relation $\beta_{out}(t)=\beta_{in}(t) + \sqrt{\kappa}C(t)$. \textit{I.e.}, after having solved Eq.(\ref{eq:cav}) for $C(t)$, we obtain the waveform of the reflected photon. Passing to the frequency domain by a Fourier transformation, Eq.(\ref{eq:cav}) becomes an algebraic equation, $-\omega \tilde{C}(\omega)=-i\frac{\kappa}{2}\tilde{C}(\omega)-i\sqrt{\kappa}\tilde{\beta}_{in}(\omega)$, and we readily find the frequency dependent reflection coefficient of the cavity,
\begin{equation} \label{eq:reflection}
R(\omega)= \tilde{\beta}_{out}(\omega)/\tilde{\beta}_{in}(\omega)= 1-\frac{\kappa}{i\omega-\kappa/2}.
\end{equation}
This coefficient is equal to unity for $|\omega| \gg \kappa$ and changes sign close to resonance $|\omega| \ll \kappa$.

\subsection{Cavity field and atomic system}

When the cavity field interacts with atoms inside the cavity, we must identify and solve the corresponding coupled equations for the amplitudes of the states occupied by the system, with Eq.(\ref{eq:cav}) providing the coupling to the input field.

\subsubsection{Hamiltonians}

The qubit atom has three states, $\ket{0_q},\ \ket{1_q} $ and $\ket{2_q}$, and we assume that prior to the arrival of the light pulse, an adiabatic or suitably tailored pulse drives the $\ket{1_q}$ qubit level into a dressed eigenstate of the Hamiltonian,
\begin{align}
  H_{\text{qub}} = \Delta\ket{2_q} \bra{2_q} + \epsilon (\ket{1_q} \bra{2_q}+\ket{2_q} \bra{1_q}) ,\nonumber
\end{align}
while the qubit state $\ket{0_q}$ is left unchanged. The laser field with detuning $\Delta$ and Rabi frequency $\epsilon$ is left on during the entire reflection process, and it is useful to introduce the dressed eigenstates,
\begin{align}\label{eq:dressedframe}
\ket{\bar 1_q}=&\cos(\Theta)\ket{1_q}-\sin(\Theta)\ket{2_q},\nonumber\\
\ket{\bar 2_q}=&\cos(\Theta)\ket{2_q}+\sin(\Theta)\ket{1_q}
\end{align}
with $\Theta=\tan^{-1}(2\epsilon/\Delta)/2$ and the energy separation given by $\bar\Delta=E_2-E_1=\sqrt{\Delta^2+4\epsilon^2}$.

Each ancilla atom, labelled here by an index $m$, couples to a classical field
with Rabi frequency $\Omega$ and to the cavity field with coupling strength $g_m$
\begin{align}
H^m_{\textrm{anc}} =&  \Omega(\ket{1^m_a} \bra{2^m_a}+h.c.)
 +i\frac{\Gamma}{2}\ket{1^m_a} \bra{1^m_a}+i\frac{\gamma}{2}\ket{2^m_a} \bra{2^m_a}\nonumber\\
H^m_{JC}=&g_m \ket{1^m_a} \bra{0^m_a}b+g_m \ket{0^m_a} \bra{1^m_a}b^\dagger,
\end{align}
where we have included damping terms, representing the decay of the excited ancilla states. Such decay will cause a loss of norm, ultimately reflected in a reduction of the output single photon field amplitude, and it will constitute part of the gate error of our protocol. Note that we did not incorporate similar decay terms in the qubit Hamiltonian. We shall treat the dressed states as being populated throughout the photon reflection process, and estimate the decay error from the excited state population in the dressed state and the duration of the gate.

Finally, we represent the Rydberg interaction between the excited qubit and an ancilla atom by the blockade strength $\mathcal{B}_m$,
\begin{align}
 H^m_{block} =&   \mathcal{B}_m   \ket{2_q 2^m_a} \bra{2_q 2^m_a},
\end{align}
where different values of $\mathcal{B}_m=C_3/{r_m}^3$ are due the different distances $r$ between the ancilla and qubit atoms. Mutual interaction between excited ancilla atoms does not occur, since the incident single photon field allows only the excitation of a single ancilla atom.

The blockade interaction can be written in the dressed state basis of the qubit
\begin{align}
 H_{block} =& \cos^2(\Theta)\mathcal{B}_m\ket{\bar 2_q 2^m_a}\bra{\bar 2_q 2^m_a}\nonumber\\
 &+\sin^2(\Theta)\mathcal{B}_m\ket{\bar 1_q 2^m_a}\bra{\bar 1_q 2^m_a}
\nonumber\\
&+\mathcal{B}_m\cos(\Theta)\sin(\Theta) \left(\ket{\bar 2_q 2_a}\bra{\bar 1_q 2_a}+\ket{\bar 2_q 2_a}\bra{\bar 1_q. 2_a}\right)\nonumber
\end{align}
Transitions into the dressed states $\ket{\bar 2_q}$ are suppressed by the energy separation $\bar{\Delta}$ between the dressed states.

\subsubsection{Schr\"odinger equation}

We solve the Schr\"odinger equation by expanding the state of the entire system excited by the incident photon wave packet on a complete basis,
\begin{align}
\ket{\psi}(t) & =  C_1(t)\ket{1_c \bar 1_q 0_a}+C_2(t)\ket{1_c \bar 2_q 0_a}\nonumber \\
& +\sum_m A_1^m(t) \ket{0_c \bar 1_q 1_a^m} + A_2^m(t) \ket{0_c \bar 2_q 1_a^m} \nonumber \\
& +\sum_m B_1^m(t) \ket{0_c \bar 1_q 2_a^m} + B_2^m(t) \ket{0_c \bar 2_q 2_a^m},
\end{align}
where the collective ancilla states with the $m^{th}$ atom excited is denoted
$\ket{1(2)^m_a}=\ket{0,0, ..,1(2), ..., 0}$.

If the qubit atom is initially in the state $\ket{1_q}$ and, thus, transferred to the dressed state $\ket{\bar 1_q}$, the Schr\"odinger equation for the amplitudes read ($\hbar = 1$)
\begin{align} \label{eq:coupled-t}
i\dot{C}_1(t)&=\sum_m g_m A_1^m(t)  -i\frac{\kappa}{2} C_1(t) - i\sqrt{\kappa} \beta_{in}(t)\nonumber \\
i\dot{C}_2(t)&=\sum_m g_m A_2^m(t)  +(\overline{\Delta} -i\frac{\kappa}{2}) C_2(t)\nonumber \\
i\dot{A}_1^m (t)&= \Omega B_1^m(t) + g_m  C_1(t) -i\frac{\Gamma}{2}A_1^m(t)\nonumber \\
i\dot{A}_2^m (t)&= \Omega B_2^m(t) + g_m  C_2(t) + (\overline{\Delta} -i\frac{\Gamma}{2})A_2^m(t)\nonumber \\
i\dot{B}_1^m (t)&= \Omega A_1^m(t) + (\sin^2(\Theta)\mathcal{B}_m-i\frac{\gamma}{2})B_1^m(t)\nonumber \\
&+ \cos(\Theta)\sin(\Theta)\mathcal{B}_m B_2^m(t)\nonumber \\
i\dot{B}_2^m (t)&= \Omega A_2^m(t) + (\cos^2(\Theta)\mathcal{B}_m+\overline{\Delta}-i\frac{\gamma}{2})B_2^m(t)\nonumber \\
&+ \cos(\Theta)\sin(\Theta)\mathcal{B}_m B_1^m(t).
\end{align}
Note that the incident field amplitude, pertaining to the initial dressed state of the qubit atom, enters as the inhomogeneous term in the first equation, similar to the driving term in Eq.(\ref{eq:cav}).

The equations (\ref{eq:coupled-t}) can be converted to algebraic equations in frequency domain by a Fourier transform,
\begin{align} \label{eq:coupled}
-\omega\tilde{C}_1(\omega)&=\sum_m g_m \tilde{A}_1^m(\omega) -i\frac{\kappa}{2} \tilde{C}_1(\omega) -i\sqrt{\kappa} \tilde{\beta}_{in}(\omega)\nonumber \\
-\omega\tilde{C}_2(\omega)&=\sum_m g_m \tilde{A}_2^m(\omega)  +(\overline{\Delta} -i\frac{\kappa}{2}) \tilde{C}_2(\omega)\nonumber \\
-\omega\tilde{A}_1^m (\omega)&= \Omega \tilde{B}_1^m(\omega) + g_m  \tilde{C}_1(\omega) -i\frac{\Gamma}{2}\tilde{A}_1^m(\omega)\nonumber \\
-\omega\tilde{A}_2^m (\omega)&= \Omega \tilde{B}_2^m(\omega) + g_m  \tilde{C}_2(\omega) + (\overline{\Delta} -i\frac{\Gamma}{2})\tilde{A}_2^m(\omega)\nonumber \\
-\omega\tilde{B}_1^m (\omega)&= \Omega \tilde{A}_1^m(\omega) + (\sin^2(\Theta)\mathcal{B}_m-i\frac{\gamma}{2})\tilde{B}_1^m(\omega)\nonumber \\
&+ \cos(\Theta)\sin(\Theta)\mathcal{B}_m \tilde{B}_2^m(\omega)\nonumber \\
-\omega\tilde{B}_2^m (\omega)&= \Omega \tilde{A}_2^m(\omega) + (\cos^2(\Theta)\mathcal{B}_m +\overline{\Delta} -i\frac{\gamma}{2})\tilde{B}_2^m(\omega)\nonumber \\
&+ \cos(\Theta)\sin(\Theta)\mathcal{B}_m \tilde{B}_1^m(\omega).
\end{align}

These equations can be readily solved in a sequence of analytical steps: First, the lower pair of equations is solved and the resulting $\tilde{B}_1^m,\ \tilde{B}_2^m$ are inserted in the middle pair of equations. This allows solution for the pair of variables $\tilde{A}_1^m,\ \tilde{A}_2^m$ in terms of $\tilde{C}_1$ and $\tilde{C}_2$. The resulting closed pair of equations for the two variables $\tilde{C}_1(\omega),\ \tilde{C}_2(\omega)$ is solved by inverting a $2\times 2$ matrix with $\omega$-dependent coefficients where we may, however, have to evaluate the sums of terms representing the coupling to different ancilla atoms numerically. Even for several thousand ancilla atoms, these sums are readily evaluated, while for much larger ensembles we have recourse to a binning of the atoms according to their interaction strengths $\mathcal{B}_m$ with the qubit atom. The sums over $m$ in the equations for $\tilde{C}_1$ and $\tilde{C}_2$ can then be evaluated as population weighted sums or integrals.

\subsubsection{Gate fidelity from reflection coefficient}

The output field is given by the solution to the set of coupled equations (\ref{eq:coupled-t}) for the single photon amplitudes $C_1(t),\ C_2(t)$. We are interested in the field amplitude, correlated with the qubit dressed state $\ket{\bar{1}_q}$,
$\tilde{\beta}_{out}(\omega)=\tilde{\beta}_{in}(\omega) - \sqrt{\kappa}\tilde{C}_1(\omega)$, while an (undesired) photon wave packet $\tilde{\xi}_{out}(\omega) = - \sqrt{\kappa}\tilde{C}_2(\omega)$ is correlated with transfer of the qubit to the dressed state $\ket{\tilde{2}_q}$.
The outcome of the calculation is the complex reflection coefficient $R_1(\omega) = \tilde{\beta}_{out}(\omega)/\tilde{\beta}_{in}(\omega)$ , which determines the modification of the photon wave packet by the reflection process and thus forms the basis for the fidelity analysis of the gate. The reflection coefficient $R_0(\omega)$ pertaining to the qubit state $\ket{0_q}$ attains the same values as in \cite{wade_single-atom_2016},
\begin{equation}
R_0(\omega)= 1-\kappa\bigl(\frac{\kappa}{2}-i\omega + \frac{G^2}{\frac{\Gamma}{2}-i\omega + |\Omega|^2/(\frac{\Gamma}{2}-i\omega)}\bigr)^{-1},
\end{equation}
where we have introduced the collectively enhanced cavity coupling, $G^2 = \sum_m |g_m|^2$. As we aim to produce a phase gate, the reflection coefficient should have a sign, $R_q(\omega)=\pm 1$, that depends on the qubit state and which is constant over the spectral components of the incident photon. We use the state-averaged gate-overlap fidelity, given by the expression \cite{pedersen_fidelity_2007}
\begin{equation} \label{eq:fidelity}
F_{at-ph} = \frac{1}{20}(\mathrm{Tr}(M M^\dagger)+|M M^\dagger|^2),
\end{equation}
with $M$ the overlap matrix between desired (\textsc{cphase}) and physically obtained evolutions.
The calculations include decay of the ancilla atoms through the damping rates $\Gamma,\ \gamma$, but decay of the qubit atom Rydberg level has so far been disregarded. The application of the Fourier transform to the equations (\ref{eq:coupled}) assume time independent coupling coefficients, and hence that the qubit atom is excited before and becomes de-excited only after the reflections process. To estimate the loss of fidelity, due to decay of the qubit atom, we assume that this occurs independently of the evolution of the ancilla atoms and the field, and that it merely amounts to the decay probability from the dressed state $\ket{\overline{1}}$ during the reflection of the field, $\eta=\exp(-\gamma T  \sin^2(\Theta)/2)$.

The reflection phase associated with the qubit states leads to an atom-photon phase gate, with the photonic qubit basis $\ket{0(1)_{ph}}$ represented by zero and one photon states, incident on the cavity. Due to the physics of the reflection process, the non-zero phase $\pi$  appears on the $\ket{0_q 1_{ph}}$ state component, and thus yields a \textsc{cphase} gate that is controlled by the $\ket{0_{q}}$ qubit state. For a desired phase rotation $\varphi$ on the $\ket{0_q 1_{ph}}$ state, Eq.(\ref{eq:fidelity}) leads to the expression for the average gate error,
\begin{eqnarray} \label{eq:fidelity}
E  =& 1-&F_{at-ph} \\
  =& 1-&\frac{1}{20}(1+|T_0|^2+\eta^2+ |T_1|^2\eta^2\nonumber\\
 &&+|1+e^{i\varphi}T_0+\eta +T_1\eta|^2),\nonumber
\end{eqnarray}
where
$T_q $=$ \int_{-\infty}^{\infty} d\omega \beta_{in}(\omega)\beta_{out}^*(\omega) $=$  \int_{-\infty}^{\infty} d\omega |\beta_{in}(\omega)|^2 R_q^*(\omega)$, are overlap integrals between the incident and reflected photon wave packets in case of the $0$ and $1$ qubit states.

In our numerical analysis we consider photon wavepackets of the form $\phi(t)=max\{0,\mathcal{A}(\exp(-(t-T/2)^2/2\sigma_T^2)-\exp(-(T)^2/8\sigma_T^2))\}$. While the frequency dependence of the reflection coefficient favors narrow bandwidth pulses, the decay of the ancilla and qubit atoms during the reflection favors short, and hence broadband, pulses.

\section{Atom-photon gate results}

\subsection{Rydberg ensemble geometry}

We assume that the ensemble of ancilla atoms is arranged in a Gaussian distribution around the central qubit atom,
\begin{equation} \label{eq:cloud}
\rho(r)dV= \rho_0\exp(-(r)^2/2R_c^2))dV,
\end{equation}
parametrized by the width parameter $R_c \simeq 10 \mu$m and the peak density $\rho_0 \simeq 10^{14}-10^{15} \mathrm{cm}^{-3}$  \cite{brennecke_cavity_2007,murch_observation_2008}.  As shown in Fig.~\ref{fig:1}, the dispersive Rydberg blockade as well as the anti-blockade case (corresponding to dressed qubit level anti-crossing) will provide mechanisms to block the EIT mechanism depending on the distance between the qubit and the ancilla atoms.  Solving Eqs.~\ref{eq:coupled} with the appropriate parameters yields the frequency dependent reflection coefficient and the fidelity of the phase gate. We first consider the case of dispersive blockade to compare to previous work using shelved Rydberg states.

 \begin{figure}
	\includegraphics[width=.47\textwidth]{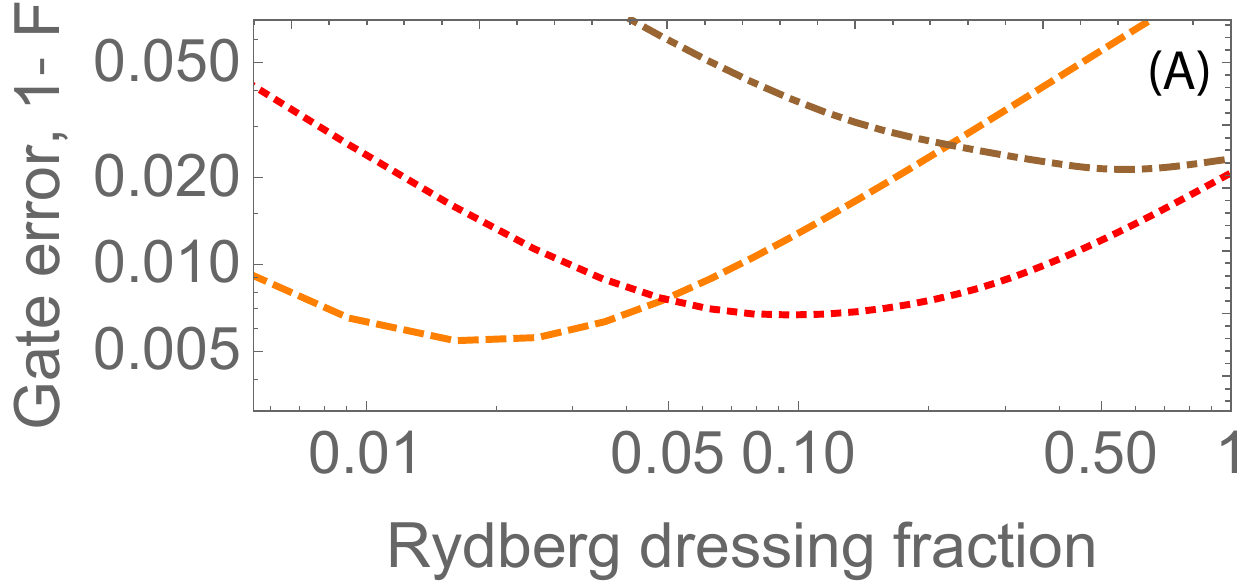}
    \includegraphics[width=.47\textwidth]{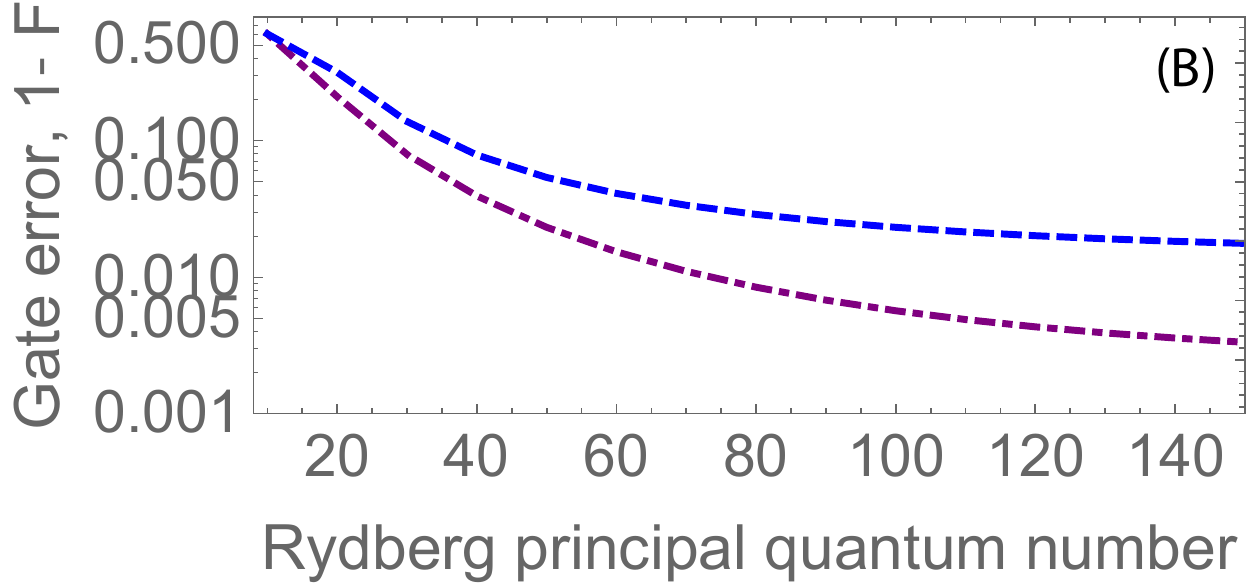}
	\caption{(A) Gate error \eqref{eq:fidelity} as a function of the qubit Rydberg excited stat populatione, for a spherical Gaussian distribution of ancillas around the central qubit atom (parameters described in the text).  The dashed orange line is for a bandwidth of 0.1MHz, the red dotted for 1MHz, and the dot-dashed brown for 10MHz. (B)  Minimum error as a function of the principal quantum number $n$ assuming approximate scaling as $n^4$ and $n^3$ of the  blockade strength and lifetime of the Rydberg state, respectively. For the purple dot-dashed line the bandwidth and Rydberg dressing fraction are chosen at an optimal point, while for the dashed blue line, only the bandwidth is optimized while the dressing fraction is fixed at 1, corresponding to the Rydberg shelving scheme. The minima of the orange and brown lines in the upper panel correspond approximately to the parameters of the purple and blue lines at $n=100$. }\label{fig:2}
\end{figure}

\subsection{Red detuned dressing}

The first main result of the paper is that detuning the qubit and operating it in the corresponding dressed state can drastically reduce decay from the excited state while still allowing significant blockade of the EIT mechanism via a large number of participating atoms.

To demonstrate this improved coupling mechanism we show how the error scales as a function of the dressing parameter $\Theta$. In Fig. 2A we see three lines corresponding to three photon bandwidths of (10MHz, 1MHz, 0.1MHz). The stated bandwidths of the pulses are calculated by Fourier transform and numerical extraction of the standard deviation.  For each pulse, the average error  (Eq.~\ref{eq:fidelity}) is minimized numerically over $\kappa>20$MHz, while other parameters are set to realistic values of $\Omega=30$MHz, $\Delta=300$MHz, $R_c=5\mu$m, $C_3=-30 \text{GHz}/\mu m^3$, $\Gamma=3$MHz, $\gamma=1$kHz, $g_m=0.5$MHz, $\rho_0=10^{13}/\mathrm{cm}^3$. $\epsilon$ is calculated for given $\Delta$ to satisfy the dressing value $\Theta$ on the x-axis. A qubit Rydberg population  greater than 0.5 corresponds to a dressed state that is mostly composed of the Rydberg state, and therefore a different initial state is assumed in our calculation (for comparison purposes).

We see that the optimal dressing scales with this bandwidth because as the pulses are made faster the coupling should be stronger to enable blocking the photon.  Nonetheless, we see that the global optimal regime is actually very low dressing strength and low bandwidth where the photon can still be fully blockaded while the excited qubit state has greatly enhanced lifetime.

In Fig.~\ref{fig:2}B, the same physical system is used, but now the Rydberg level rather than the dressing is varied.  The same parameters are used but the optimal value of dressing (minimum in Fig.~\ref{fig:2}A) is assumed.  We apply an approximate scaling where $C_3\sim-300 n^4\text{Hz}/\mu m^3$ and $\gamma \sim n^{-3}$GHz.  As expected from this scaling, the dressing is most important in the low error regime, where less Rydberg population and longer pulses can be used.

\subsection{Blue detuned dressing}

For the situation where the blockade shift and Rydberg detuning have opposite signs, as in Fig.~\ref{fig:1}D, the system has a resonant excitation pathway to the fully Rydberg excited qubit state and a Rydberg excited ancilla atom at a distance obeying $\mathcal{B}_m = C_3/{r_m}^3 \simeq \Delta$. The coupling to this state breaks the EIT mechanism and splits the dark state by $\pm \epsilon$, causing reflection of the incident photon. The reflection coefficient $R_1(\omega)$ is given by $\tilde{C}_1(\omega)$, which is found by solving the full set of equations (15). The first line in Eq.(15), however, indicates how much individual ancilla atoms contribute to the reflection coefficient,
%
\begin{equation} \label{eq:reflection2}
R_1(\omega)= \frac{\tilde{\beta}_{out}(\omega)}{\tilde{\beta}_{in}(\omega)}= 1-\frac{\kappa}{i\omega-\kappa/2+\sum_m g_m\tilde A_1^m(\omega)/\tilde C_1(\omega)}.
\end{equation}
Only ancilla atoms in a thin shell obey the two-atom resonance condition, and their large values of $\tilde{A}_1^m(\omega)/\tilde{C}_1(\omega)$, shown in Fig. 3B, cause $R_1(\omega)$ to change from $-1$ to 1.

%

\begin{figure}
    \includegraphics[width=.67\textwidth]{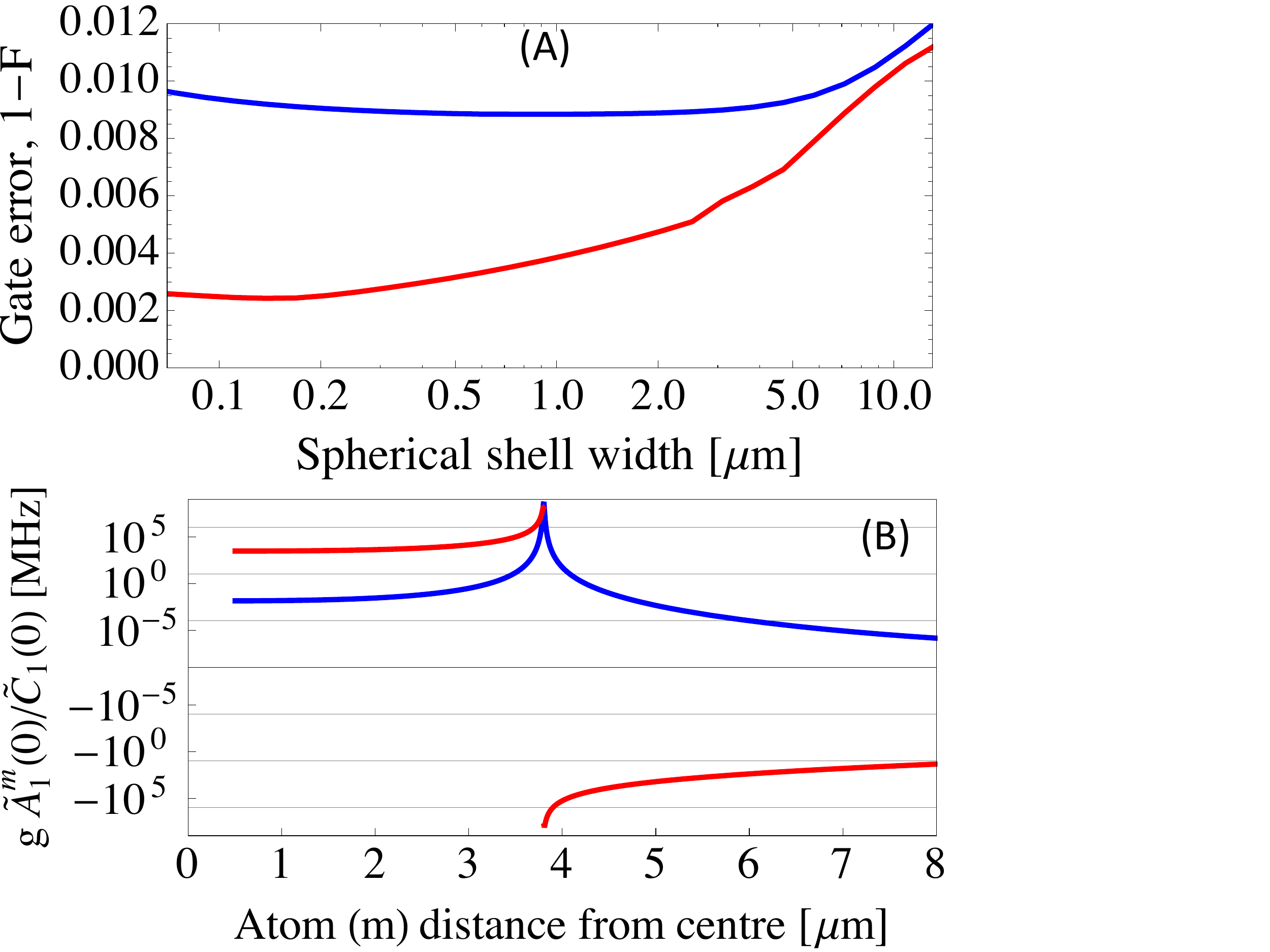}
	
	\caption{(A)  Gate error \eqref{eq:fidelity} as a function of the width of a shell of ancilla atoms included from a Gaussian distribution \eqref{eq:cloud} at about 5$\mu$m from the qubit. Blue upper 
    and red lower lines are for blue and red detuning of the dressing laser, respectively. The coupling strength and dressed laser detuning parameters are optimized as in Fig.~\ref{fig:2}, with other parameter values discussed in text. (B) The real and imaginary parts of the quantity $g \tilde A^m_1(\omega=0)/\tilde C_1(\omega=0)$, is plotted as a function of radial distance from the qubit. A peak occurs at the distance where the resonance in Fig.~\ref{fig:1}D occurs according to the dressing laser detuning. \label{fig:3}}
\end{figure}

The gate error is plotted in Fig.~\ref{fig:3}A, as a function of the width of the distribution of atoms included from the Gaussian distribution \eqref{eq:cloud} around the resonant distance. In our calculations we assume the physical parameters $\Omega=30$MHz, $R_c=10\mu$m, $C_3=-18 \text{GHz}/\mu m^3$, $\Gamma=3$MHz, $\gamma=1$kHz, $g_m=0.5$MHz, $\rho_0=10^{13}/\mathrm{cm}^3$, $\Theta=0.15$, and an incident photon bandwidth 0.1MHz.  Allowing the Rydberg state population ($\cos(\Theta)$) and the photon bandwidth to vary does not significantly improve the fidelity. We see that the resonance case outperforms its dispersive version as well as the Rydberg shelving approach and reaches errors in the few $10^{-3}$ range, satisfying fault-tolerance thresholds for several error correcting codes.
Compared to the strong coupling to single atoms \cite{reiserer_quantum_2014}, the errors here are close two orders of magnitude lower.

Further improvement in the gate fidelity may be obtained by spatial or spectral tailoring of the ancilla coupling parameters. Rather than relying on the $C_3/r_m^3$ dependence one may, e.g., exploit long range local minima in more complex energy spectra \cite{greene_creation_2000,boisseau_macrodimers:_2002} to increase the number of ancilla atoms experiencing the two-atom resonance.

\subsection{Arbitrary controlled-phase}

In quantum computing, a key requirement is a universal gate set that spans all possible operations. Our {\sc cphase} gate in combination with local qubit operations provide such a gate set, but being able to achieve other controlled phases than $\pi$ can be of significance in the efficient composition of more complicated operations. As illustrated by the empty cavity (9), the reflection phase varies between 0 and $\pi$ as the frequency of the photon approaches cavity resonance. As the complex phase of our reflection coefficients $R_0(\omega)$ and $R_1(\omega)$ have different frequency dependences, we may hence investigate the ability to generate arbitrary controlled phases.

\begin{figure}
    \includegraphics[width=.47\textwidth]{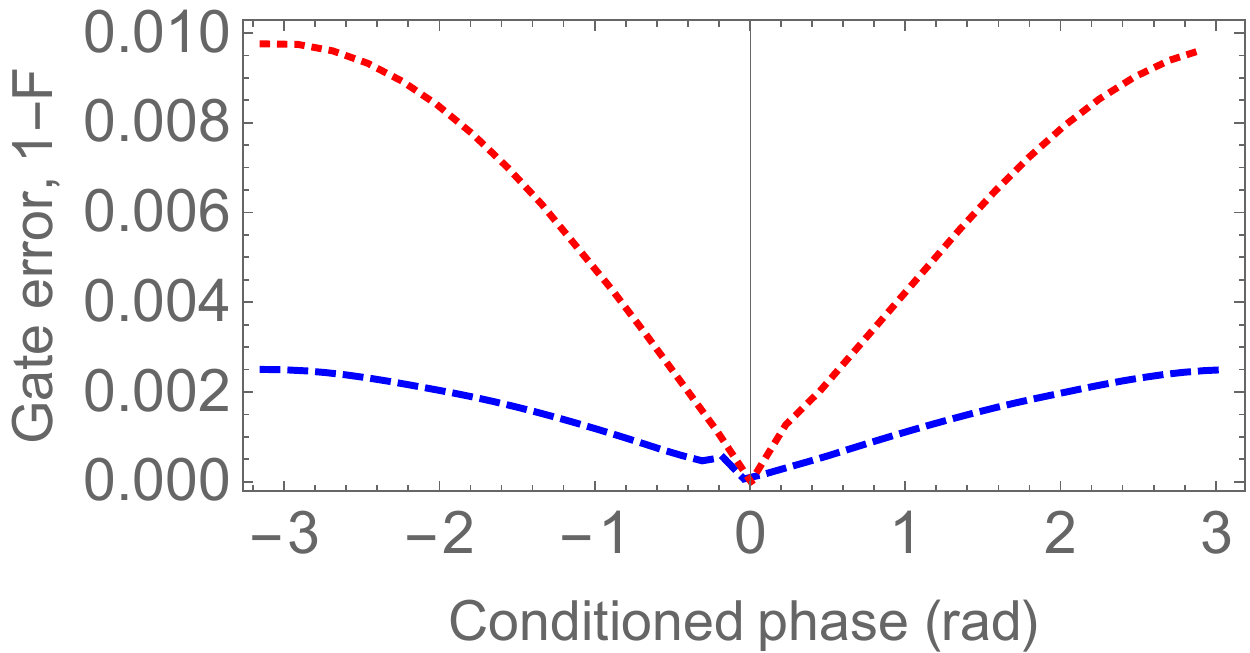}
	 \caption{Gate error \eqref{eq:fidelity} as a function of the desired phase acquired on the $\ket{0_q 1_{ph}}$ state vs the other 3 qubit product states.  The red dashed line corresponds to the full cloud geometry with red-detuned dressing, while the blue is for a shell of width 1$\mu$m at a distance of 5$\mu$m for blue-detuned dressing. Other parameters are the same as in Fig.~\ref{fig:3}.
     \label{fig:4}}
\end{figure}

In Fig.~\ref{fig:4}, we plot the error associated with achieving different phases $\varphi$ in Eq.~\ref{eq:fidelity}.  The simulation parameters are the same as for Fig.~\ref{fig:3} with a shell width taken as $1\mu$m and the central freqeuncy of the incoming photon pulse relative to the cavity mode being the crucial optimization parameter.
The most difficult to achieve phase is not surprisingly the largest phase difference $\pi$ between the reflection coefficients, which is the phase assumed in the rest of the figures.

\section{Implementation with superconducting circuits}

The use of an ancilla system as mediator of the interaction between a stationary and a flying qubit is not restricted to atomic systems. Another prominent candidate for quantum information technologies is the circuit-QED architecture, that may ultimately rely on effective microwave communication between separate chips with superconducting qubits for highly parallelized processing within a single or multiple dilution refrigerators.

\begin{figure}
	\includegraphics[width=.50
    \textwidth]{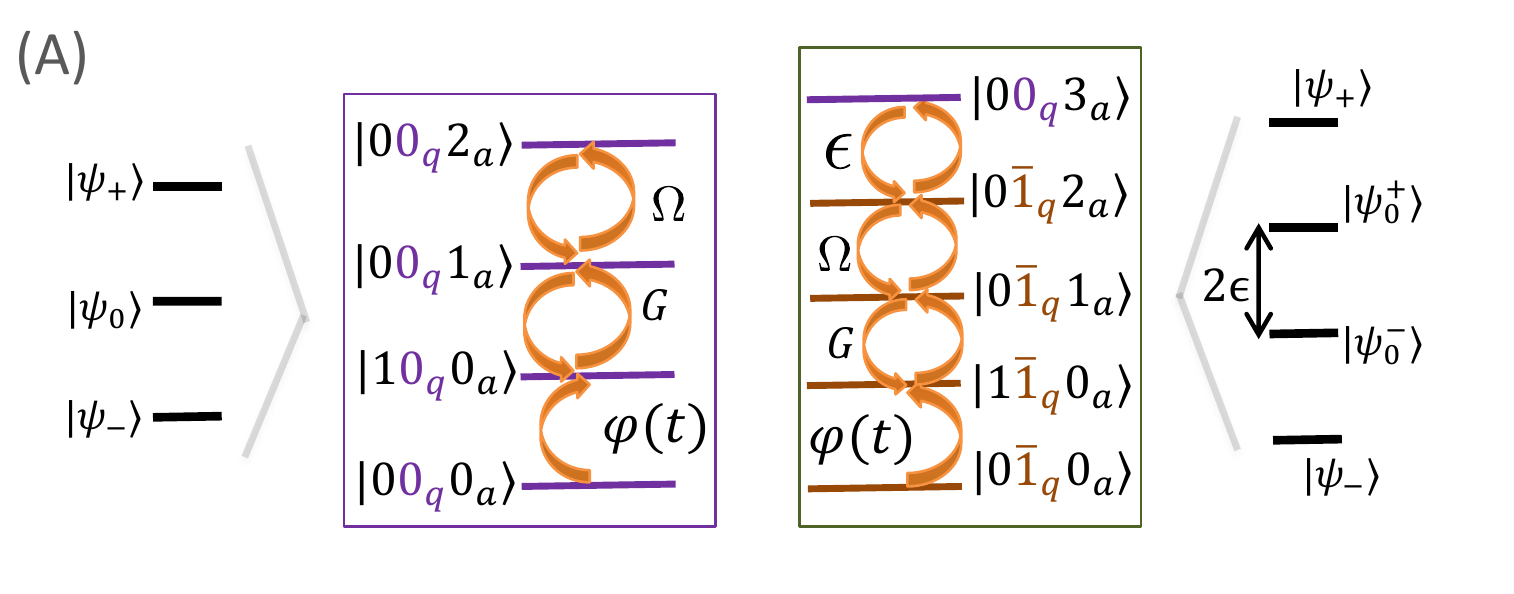}
    \includegraphics[width=.48\textwidth]{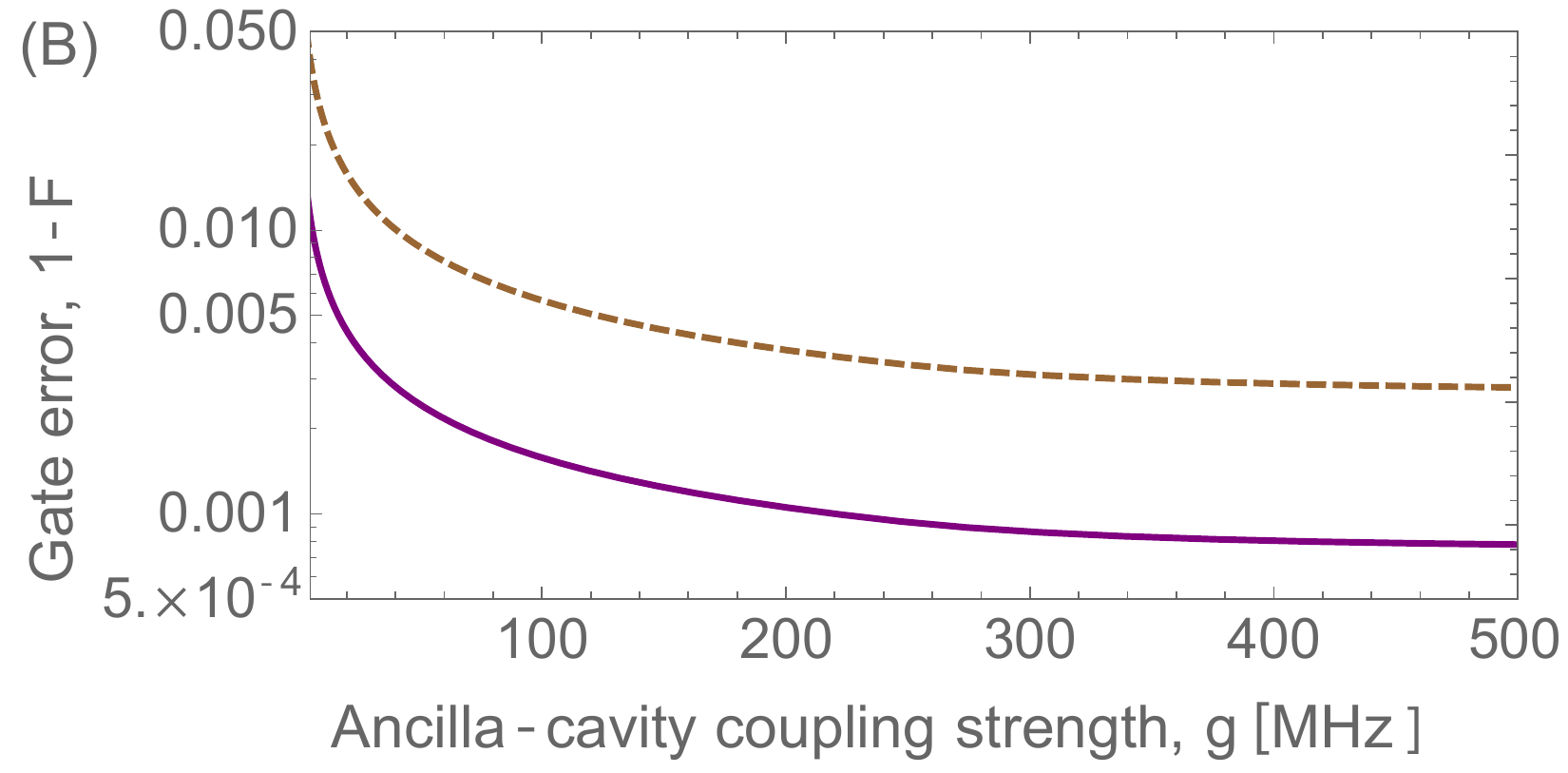}
	 \caption{(A) energy level diagram and couplings for implementing the protocol with superconducting transmon qubits and microwave cavity. (B) Gate error \eqref{eq:fidelity} for the circuit-QED setup from Fig.~\ref{fig:1}B as a function of the coupling from the cavity to the ancilla transmon qubit. The dotted line is for a lifetime of the qubit of 5$\mu$s, while the solid purple is for 33$\mu$s. The ancilla lifetime is assumed to be half the register qubit's. Other parameters are described in the text.
     }\label{fig:5}
\end{figure}

The arrangement is depicted in Fig.~\ref{fig:1}B, 
corresponding to a microwave transmission line cavity with two superconducting transmon. The left qubit transmon is restricted to its two lowest energy eigenstates $\ket{0(1)_q}$ while the right ancilla transmon is restricted to its four lowest levels $\ket{i_a}$, $i=0 .. 3$.
As in the atomic case, we assume that the incident microwave photon is resonant with the resonator, which is in turn resonantly coupled with strength $g$  with the $\ket{0_a} \leftrightarrow \ket{1_a}$ ancilla transmon transition, and we assume a classical microwave drive is applied with strength $\Omega$ to the $\ket{1_a} \leftrightarrow \ket{2_a}$ ancilla transmon transition. In addition we assume degeneracy of the product states $\ket{2_a1_q}$ and $\ket{3_a0_q}$ and a coupling between them with strength $\varepsilon$. This energy structure is depicted in Fig.~\ref{fig:5}A.

In the limit of low charge dispersion, the transmon qubits can be accurately modeled as Duffing oscillators with low anharmonicity \cite{khani_optimal_2009,khani_high-fidelity_2012}, and with Hamiltonian
\begin{eqnarray}
H&=&\omega a^\dagger a + \omega_1 b^\dagger b + \omega_2 c^\dagger c + \alpha_1 (b^\dagger b)^2 + \alpha_2 (c^\dagger c)^2 \nonumber\\
&&+ g(b^\dagger a + a^\dagger b)+ \epsilon(b^\dagger c + c^\dagger b)+\frac{\Omega}{\sqrt 2}(b^\dagger+ b)\nonumber
\end{eqnarray}
The rotating frame Hamiltonian where we keep only the relevant levels then reads
\begin{eqnarray} \label{eq:SCeit}
H&=&g(\ket{1}\bra{0}_a a^\dagger + a \ket{0}\bra{1}_a)+\Omega(\ket{1}\bra{2}_a+\ket{2}\bra{1}_a) \nonumber\\
&&+\epsilon(\ket{2}\bra{3}_a \ket{1}\bra{0}_q + \ket{3}\bra{2}_a \ket{0}\bra{1}_q).
\end{eqnarray}
where $\Gamma_i$ is the decay from the $i$-th level of the ancilla atom.

We see from Fig.~\ref{fig:5}A that, for the microwave photon incident on the initial transmon qubit $\ket{0_q}$ state, the same EIT configuration (left three level ladder) appears as in the previous section, while for the qubit $\ket{1_q}$ state, the coupling to the state $\ket{2_a1_q}$, splits the upper level, and effectively couples the cavity resonantly to the two lowest state of the ancilla transmon. Note that EIT has been previously studied \cite{abdumalikov_electromagnetically_2010,murali_probing_2004} previously for superconducting qubits, but here we show can it can be applied for a high-fidelity quantum switch.

Solving the Schroedinger equation for the different state amplitudes in the input-output theory yields the complex reflection coefficients,
\begin{align}
R_{q}(\omega)= 1{}&{} - \kappa \left\{\frac{\kappa}{2} - i \omega +   |g|^2 \left[ \frac{\Gamma_{1}}{2}-i\omega \phantom{+\frac{|\Omega|^2}{\frac{\Gamma_{r}}{2} - i \omega + \frac{|V_{jn}|^2}{\frac{\Gamma_p}{2}+i(\delta-\omega)}}}\right .\right . \nonumber \\
&\left.\left.\qquad\qquad{}+\frac{2|\Omega|^2}{\frac{\Gamma_{2}}{2} - i \omega + q\frac{ 3|\epsilon|^2}{\frac{\Gamma_3}{2}-i\omega}}\right]^{-1}\right\}^{-1},\label{eq:full_R}
\end{align}
which is effectively controlled by the qubit state if $\varepsilon > \Omega \gg \Gamma_i,\kappa$
Note that we must also assume sufficient anharmonicity of the transmon energy levels to avoid excitation of the qubit transmon to higher states \cite{theis_simultaneous_2016,goerz_charting_2017} and undesired couplings among ancilla levels due to the strong classical microwave drive $\Omega$.
This puts requirements on the  anharmonicities $\alpha_i \gg \Omega,\epsilon$ of the transmons. These requirements can be fulfilled in experiments, where we may have transmon state lifetimes of tens of microseconds and couplings $\varepsilon$ and $\Omega$ in the tens to low hundreds of MHz can be obtained, while anharmonicities are typically in the few hundreds of MHz. This setup also corresponds relatively well to the parameter regimes we have chosen for the neutral atom Rydberg systems, where Rabi frequencies can be significantly lower but hyperfine and Rydberg lifetimes significantly longer.

We use the same theoretical expression for state overlaps and fidelities as in the previous section to obtain the {\sc cphase} gate error, and we plot this quantity as a function of the cavity vacuum Rabi frequency coupling to the ancilla transmon in  Fig.~\ref{fig:5}B.  The dashed brown curve is simulated with a qubit lifetime of $5\mu$s while the solid purple lines is for $33\mu$s. The higher transmon level lifetimes decrease with Fock number. The coupling between transmons is taken as $\epsilon=150$MHz and the other parameters are optimized numerically as before, specifically the bandwidth and EIT laser strength, which are in the 10MHz and 100MHz range, respectively, but increase with $g$. Error rates are limited to interaction times of hundreds of $ns$, similar to circuit-QED two-qubit gates which have similar limitations. Thus, qubit-photon gates can achieve similar intrinsic fidelities as qubit-qubit gates \cite{goerz_charting_2017}, with photon traveling losses likely to be the most severe bottleneck for scalable technology in the foreseeable future.

\section{conclusion}

We have shown that it is possible to strongly couple single photons to single stationary qubits by means of ancilla systems, which may be optimized for communication while the qubit may be optimized for storage or, e.g.,  for local interaction within a register of qubits. We presented quantitative analyses, improving on earlier proposals for neutral qubit atoms interacting with an ancilla medium of atoms by strong dipolar forces among Rydberg excited states and we presented an original implementation for superconducting circuits with engineered resonant interactions between the qubit and ancilla transmons within wave guide resonators.
For both implementations, the proposal meets a need to achieve high-performance or fault-tolerant scalable quantum operations, and we believe that the main ideas may be applied to other systems, e.g, hybrid atom and solid state systems, where a mediator is needed for the coupling to, and  communication by, phonons, plasmons, spin waves and other extended quantum degrees of freedom.

\section{acknowledgements}
This work was supported by the ARL-CDQI program
through cooperative agreement W911NF-15-2- 0061, and the Villum Foundation. We thank L.~Buchmann,  D.~Petrosyan and M.~Saffman for fruitful discussions.

\bibliography{Zotero}

\begin{thebibliography}{44}
\expandafter\ifx\csname natexlab\endcsname\relax\def\natexlab#1{#1}\fi
\expandafter\ifx\csname bibnamefont\endcsname\relax
  \def\bibnamefont#1{#1}\fi
\expandafter\ifx\csname bibfnamefont\endcsname\relax
  \def\bibfnamefont#1{#1}\fi
\expandafter\ifx\csname citenamefont\endcsname\relax
  \def\citenamefont#1{#1}\fi
\expandafter\ifx\csname url\endcsname\relax
  \def\url#1{\texttt{#1}}\fi
\expandafter\ifx\csname urlprefix\endcsname\relax\def\urlprefix{URL }\fi
\providecommand{\bibinfo}[2]{#2}
\providecommand{\eprint}[2][]{\url{#2}}

\bibitem[{\citenamefont{Kimble}(2008)}]{kimble_quantum_2008}
\bibinfo{author}{\bibfnamefont{H.~J.} \bibnamefont{Kimble}},
  \bibinfo{journal}{Nature} \textbf{\bibinfo{volume}{453}},
  \bibinfo{pages}{1023} (\bibinfo{year}{2008}).

\bibitem[{\citenamefont{Duan and Kimble}(2004)}]{duan_scalable_2004}
\bibinfo{author}{\bibfnamefont{L.~M.} \bibnamefont{Duan}} \bibnamefont{and}
  \bibinfo{author}{\bibfnamefont{H.~J.} \bibnamefont{Kimble}},
  \bibinfo{journal}{Physical review letters} \textbf{\bibinfo{volume}{92}},
  \bibinfo{pages}{127902} (\bibinfo{year}{2004}).

\bibitem[{\citenamefont{Reiserer et~al.}(2013)\citenamefont{Reiserer, Ritter,
  and Rempe}}]{reiserer_nondestructive_2013}
\bibinfo{author}{\bibfnamefont{A.}~\bibnamefont{Reiserer}},
  \bibinfo{author}{\bibfnamefont{S.}~\bibnamefont{Ritter}}, \bibnamefont{and}
  \bibinfo{author}{\bibfnamefont{G.}~\bibnamefont{Rempe}},
  \bibinfo{journal}{Science} \textbf{\bibinfo{volume}{342}},
  \bibinfo{pages}{1349} (\bibinfo{year}{2013}).

\bibitem[{\citenamefont{Reiserer et~al.}(2014)\citenamefont{Reiserer, Kalb,
  Rempe, and Ritter}}]{reiserer_quantum_2014}
\bibinfo{author}{\bibfnamefont{A.}~\bibnamefont{Reiserer}},
  \bibinfo{author}{\bibfnamefont{N.}~\bibnamefont{Kalb}},
  \bibinfo{author}{\bibfnamefont{G.}~\bibnamefont{Rempe}}, \bibnamefont{and}
  \bibinfo{author}{\bibfnamefont{S.}~\bibnamefont{Ritter}},
  \bibinfo{journal}{Nature} \textbf{\bibinfo{volume}{508}},
  \bibinfo{pages}{237} (\bibinfo{year}{2014}), ISSN \bibinfo{issn}{0028-0836,
  1476-4687},
  \urlprefix\url{http://www.nature.com/doifinder/10.1038/nature13177}.

\bibitem[{\citenamefont{Hao et~al.}(2015)\citenamefont{Hao, Lin, Xia, Lin, Niu,
  and Gong}}]{hao_quantum_2015}
\bibinfo{author}{\bibfnamefont{Y.~M.} \bibnamefont{Hao}},
  \bibinfo{author}{\bibfnamefont{G.~W.} \bibnamefont{Lin}},
  \bibinfo{author}{\bibfnamefont{K.}~\bibnamefont{Xia}},
  \bibinfo{author}{\bibfnamefont{X.~M.} \bibnamefont{Lin}},
  \bibinfo{author}{\bibfnamefont{Y.~P.} \bibnamefont{Niu}}, \bibnamefont{and}
  \bibinfo{author}{\bibfnamefont{S.~Q.} \bibnamefont{Gong}},
  \bibinfo{journal}{Scientific reports} \textbf{\bibinfo{volume}{5}},
  \bibinfo{pages}{10005} (\bibinfo{year}{2015}).

\bibitem[{\citenamefont{Motzoi et~al.}(2016)\citenamefont{Motzoi, Halperin,
  Wang, Whaley, and Schirmer}}]{motzoi_backaction-driven_2016}
\bibinfo{author}{\bibfnamefont{F.}~\bibnamefont{Motzoi}},
  \bibinfo{author}{\bibfnamefont{E.}~\bibnamefont{Halperin}},
  \bibinfo{author}{\bibfnamefont{X.}~\bibnamefont{Wang}},
  \bibinfo{author}{\bibfnamefont{K.~B.} \bibnamefont{Whaley}},
  \bibnamefont{and} \bibinfo{author}{\bibfnamefont{S.}~\bibnamefont{Schirmer}},
  \bibinfo{journal}{Phys. Rev. A} \textbf{\bibinfo{volume}{94}},
  \bibinfo{pages}{032313} (\bibinfo{year}{2016}),
  \urlprefix\url{https://link.aps.org/doi/10.1103/PhysRevA.94.032313}.

\bibitem[{\citenamefont{Das et~al.}(2016)\citenamefont{Das, Grankin, Iakoupov,
  Brion, Borregaard, Boddeda, Usmani, Ourjoumtsev, Grangier, and
  Sørensen}}]{das_photonic_2016}
\bibinfo{author}{\bibfnamefont{S.}~\bibnamefont{Das}},
  \bibinfo{author}{\bibfnamefont{A.}~\bibnamefont{Grankin}},
  \bibinfo{author}{\bibfnamefont{I.}~\bibnamefont{Iakoupov}},
  \bibinfo{author}{\bibfnamefont{E.}~\bibnamefont{Brion}},
  \bibinfo{author}{\bibfnamefont{J.}~\bibnamefont{Borregaard}},
  \bibinfo{author}{\bibfnamefont{R.}~\bibnamefont{Boddeda}},
  \bibinfo{author}{\bibfnamefont{I.}~\bibnamefont{Usmani}},
  \bibinfo{author}{\bibfnamefont{A.}~\bibnamefont{Ourjoumtsev}},
  \bibinfo{author}{\bibfnamefont{P.}~\bibnamefont{Grangier}}, \bibnamefont{and}
  \bibinfo{author}{\bibfnamefont{A.~S.} \bibnamefont{Sørensen}},
  \bibinfo{journal}{Physical Review A} \textbf{\bibinfo{volume}{93}},
  \bibinfo{pages}{040303} (\bibinfo{year}{2016}).

\bibitem[{\citenamefont{Firstenberg et~al.}(2016)\citenamefont{Firstenberg,
  Adams, and Hofferberth}}]{firstenberg_nonlinear_2016}
\bibinfo{author}{\bibfnamefont{O.}~\bibnamefont{Firstenberg}},
  \bibinfo{author}{\bibfnamefont{C.~S.} \bibnamefont{Adams}}, \bibnamefont{and}
  \bibinfo{author}{\bibfnamefont{S.}~\bibnamefont{Hofferberth}},
  \bibinfo{journal}{Journal of Physics B: Atomic, Molecular and Optical
  Physics} \textbf{\bibinfo{volume}{49}}, \bibinfo{pages}{152003}
  (\bibinfo{year}{2016}), ISSN \bibinfo{issn}{0953-4075, 1361-6455},
  \urlprefix\url{http://stacks.iop.org/0953-4075/49/i=15/a=152003?key=crossref.c5970f07d580ff994fcb28d3ed09f4e3}.

\bibitem[{\citenamefont{Cabrillo et~al.}(1999)\citenamefont{Cabrillo, Cirac,
  García-Fernández, and Zoller}}]{cabrillo_creation_1999-1}
\bibinfo{author}{\bibfnamefont{C.}~\bibnamefont{Cabrillo}},
  \bibinfo{author}{\bibfnamefont{J.~I.} \bibnamefont{Cirac}},
  \bibinfo{author}{\bibfnamefont{P.}~\bibnamefont{García-Fernández}},
  \bibnamefont{and} \bibinfo{author}{\bibfnamefont{P.}~\bibnamefont{Zoller}},
  \bibinfo{journal}{Phys. Rev. A} \textbf{\bibinfo{volume}{59}},
  \bibinfo{pages}{1025} (\bibinfo{year}{1999}),
  \urlprefix\url{http://link.aps.org/doi/10.1103/PhysRevA.59.1025}.

\bibitem[{\citenamefont{Moehring et~al.}(2007)\citenamefont{Moehring, Maunz,
  Olmschenk, Younge, Matsukevich, Duan, and
  Monroe}}]{moehring_entanglement_2007}
\bibinfo{author}{\bibfnamefont{D.~L.} \bibnamefont{Moehring}},
  \bibinfo{author}{\bibfnamefont{P.}~\bibnamefont{Maunz}},
  \bibinfo{author}{\bibfnamefont{S.}~\bibnamefont{Olmschenk}},
  \bibinfo{author}{\bibfnamefont{K.~C.} \bibnamefont{Younge}},
  \bibinfo{author}{\bibfnamefont{D.~N.} \bibnamefont{Matsukevich}},
  \bibinfo{author}{\bibfnamefont{L.-M.} \bibnamefont{Duan}}, \bibnamefont{and}
  \bibinfo{author}{\bibfnamefont{C.}~\bibnamefont{Monroe}},
  \bibinfo{journal}{Nature} \textbf{\bibinfo{volume}{449}}, \bibinfo{pages}{68}
  (\bibinfo{year}{2007}), ISSN \bibinfo{issn}{0028-0836, 1476-4687},
  \urlprefix\url{http://www.nature.com/doifinder/10.1038/nature06118}.

\bibitem[{\citenamefont{Hensen et~al.}(2015)\citenamefont{Hensen, Bernien,
  Dreau, Reiserer, Kalb, Blok, Ruitenberg, Vermeulen, Schouten, Abellan
  et~al.}}]{hensen_loophole-free_2015}
\bibinfo{author}{\bibfnamefont{B.}~\bibnamefont{Hensen}},
  \bibinfo{author}{\bibfnamefont{H.}~\bibnamefont{Bernien}},
  \bibinfo{author}{\bibfnamefont{A.~E.} \bibnamefont{Dreau}},
  \bibinfo{author}{\bibfnamefont{A.}~\bibnamefont{Reiserer}},
  \bibinfo{author}{\bibfnamefont{N.}~\bibnamefont{Kalb}},
  \bibinfo{author}{\bibfnamefont{M.~S.} \bibnamefont{Blok}},
  \bibinfo{author}{\bibfnamefont{J.}~\bibnamefont{Ruitenberg}},
  \bibinfo{author}{\bibfnamefont{R.~F.~L.} \bibnamefont{Vermeulen}},
  \bibinfo{author}{\bibfnamefont{R.~N.} \bibnamefont{Schouten}},
  \bibinfo{author}{\bibfnamefont{C.}~\bibnamefont{Abellan}},
  \bibnamefont{et~al.}, \bibinfo{journal}{Nature}
  \textbf{\bibinfo{volume}{526}}, \bibinfo{pages}{682} (\bibinfo{year}{2015}),
  \urlprefix\url{http://dx.doi.org/10.1038/nature15759}.

\bibitem[{\citenamefont{Roch et~al.}(2014)\citenamefont{Roch, Schwartz, Motzoi,
  Macklin, Vijay, Eddins, Korotkov, Whaley, Sarovar, and
  Siddiqi}}]{roch_observation_2014}
\bibinfo{author}{\bibfnamefont{N.}~\bibnamefont{Roch}},
  \bibinfo{author}{\bibfnamefont{M.~E.} \bibnamefont{Schwartz}},
  \bibinfo{author}{\bibfnamefont{F.}~\bibnamefont{Motzoi}},
  \bibinfo{author}{\bibfnamefont{C.}~\bibnamefont{Macklin}},
  \bibinfo{author}{\bibfnamefont{R.}~\bibnamefont{Vijay}},
  \bibinfo{author}{\bibfnamefont{A.~W.} \bibnamefont{Eddins}},
  \bibinfo{author}{\bibfnamefont{A.~N.} \bibnamefont{Korotkov}},
  \bibinfo{author}{\bibfnamefont{K.~B.} \bibnamefont{Whaley}},
  \bibinfo{author}{\bibfnamefont{M.}~\bibnamefont{Sarovar}}, \bibnamefont{and}
  \bibinfo{author}{\bibfnamefont{I.}~\bibnamefont{Siddiqi}},
  \bibinfo{journal}{Phys. Rev. Lett.} \textbf{\bibinfo{volume}{112}},
  \bibinfo{pages}{170501} (\bibinfo{year}{2014}).

\bibitem[{\citenamefont{Motzoi et~al.}(2015)\citenamefont{Motzoi, Whaley, and
  Sarovar}}]{motzoi_continuous_2015}
\bibinfo{author}{\bibfnamefont{F.}~\bibnamefont{Motzoi}},
  \bibinfo{author}{\bibfnamefont{K.~B.} \bibnamefont{Whaley}},
  \bibnamefont{and} \bibinfo{author}{\bibfnamefont{M.}~\bibnamefont{Sarovar}},
  \bibinfo{journal}{arXiv.org}  (\bibinfo{year}{2015}).

\bibitem[{\citenamefont{Martin et~al.}(2015)\citenamefont{Martin, Motzoi, Li,
  Sarovar, and Whaley}}]{martin_deterministic_2015}
\bibinfo{author}{\bibfnamefont{L.}~\bibnamefont{Martin}},
  \bibinfo{author}{\bibfnamefont{F.}~\bibnamefont{Motzoi}},
  \bibinfo{author}{\bibfnamefont{H.}~\bibnamefont{Li}},
  \bibinfo{author}{\bibfnamefont{M.}~\bibnamefont{Sarovar}}, \bibnamefont{and}
  \bibinfo{author}{\bibfnamefont{K.~B.} \bibnamefont{Whaley}},
  \bibinfo{journal}{Phys. Rev. A} \textbf{\bibinfo{volume}{92}},
  \bibinfo{pages}{062321} (\bibinfo{year}{2015}),
  \urlprefix\url{http://link.aps.org/doi/10.1103/PhysRevA.92.062321}.

\bibitem[{\citenamefont{Duan et~al.}(2000)\citenamefont{Duan, Cirac, Zoller,
  and Polzik}}]{duan_quantum_2000}
\bibinfo{author}{\bibfnamefont{L.-M.} \bibnamefont{Duan}},
  \bibinfo{author}{\bibfnamefont{J.~I.} \bibnamefont{Cirac}},
  \bibinfo{author}{\bibfnamefont{P.}~\bibnamefont{Zoller}}, \bibnamefont{and}
  \bibinfo{author}{\bibfnamefont{E.~S.} \bibnamefont{Polzik}},
  \bibinfo{journal}{Physical Review Letters} \textbf{\bibinfo{volume}{85}},
  \bibinfo{pages}{5643} (\bibinfo{year}{2000}), ISSN \bibinfo{issn}{0031-9007,
  1079-7114},
  \urlprefix\url{https://link.aps.org/doi/10.1103/PhysRevLett.85.5643}.

\bibitem[{\citenamefont{Duan et~al.}(2001)\citenamefont{Duan, Lukin, Cirac, and
  Zoller}}]{duan_long-distance_2001}
\bibinfo{author}{\bibfnamefont{L.-M.} \bibnamefont{Duan}},
  \bibinfo{author}{\bibfnamefont{M.~D.} \bibnamefont{Lukin}},
  \bibinfo{author}{\bibfnamefont{J.~I.} \bibnamefont{Cirac}}, \bibnamefont{and}
  \bibinfo{author}{\bibfnamefont{P.}~\bibnamefont{Zoller}},
  \bibinfo{journal}{Nature} \textbf{\bibinfo{volume}{414}},
  \bibinfo{pages}{413} (\bibinfo{year}{2001}), ISSN \bibinfo{issn}{0028-0836},
  \urlprefix\url{http://www.nature.com/doifinder/10.1038/35106500}.

\bibitem[{\citenamefont{Pellizzari}(1997)}]{pellizzari_quantum_1997}
\bibinfo{author}{\bibfnamefont{T.}~\bibnamefont{Pellizzari}},
  \bibinfo{journal}{Physical Review Letters} \textbf{\bibinfo{volume}{79}},
  \bibinfo{pages}{5242} (\bibinfo{year}{1997}), ISSN \bibinfo{issn}{0031-9007,
  1079-7114},
  \urlprefix\url{https://link.aps.org/doi/10.1103/PhysRevLett.79.5242}.

\bibitem[{\citenamefont{Wilk et~al.}(2007)\citenamefont{Wilk, Webster, Kuhn,
  and Rempe}}]{wilk_single-atom_2007}
\bibinfo{author}{\bibfnamefont{T.}~\bibnamefont{Wilk}},
  \bibinfo{author}{\bibfnamefont{S.~C.} \bibnamefont{Webster}},
  \bibinfo{author}{\bibfnamefont{A.}~\bibnamefont{Kuhn}}, \bibnamefont{and}
  \bibinfo{author}{\bibfnamefont{G.}~\bibnamefont{Rempe}},
  \bibinfo{journal}{Science} \textbf{\bibinfo{volume}{317}},
  \bibinfo{pages}{488} (\bibinfo{year}{2007}), ISSN \bibinfo{issn}{0036-8075,
  1095-9203},
  \urlprefix\url{http://www.sciencemag.org/cgi/doi/10.1126/science.1143835}.

\bibitem[{\citenamefont{Ates et~al.}(2011)\citenamefont{Ates, Sevinçli, and
  Pohl}}]{ates_electromagnetically_2011}
\bibinfo{author}{\bibfnamefont{C.}~\bibnamefont{Ates}},
  \bibinfo{author}{\bibfnamefont{S.}~\bibnamefont{Sevinçli}},
  \bibnamefont{and} \bibinfo{author}{\bibfnamefont{T.}~\bibnamefont{Pohl}},
  \bibinfo{journal}{Physical Review A} \textbf{\bibinfo{volume}{83}}
  (\bibinfo{year}{2011}), ISSN \bibinfo{issn}{1050-2947, 1094-1622},
  \urlprefix\url{https://link.aps.org/doi/10.1103/PhysRevA.83.041802}.

\bibitem[{\citenamefont{Chang et~al.}(2014)\citenamefont{Chang, Vuletić, and
  Lukin}}]{chang_quantum_2014}
\bibinfo{author}{\bibfnamefont{D.~E.} \bibnamefont{Chang}},
  \bibinfo{author}{\bibfnamefont{V.}~\bibnamefont{Vuletić}}, \bibnamefont{and}
  \bibinfo{author}{\bibfnamefont{M.~D.} \bibnamefont{Lukin}},
  \bibinfo{journal}{Nature Photonics} \textbf{\bibinfo{volume}{8}},
  \bibinfo{pages}{685} (\bibinfo{year}{2014}), ISSN \bibinfo{issn}{1749-4885,
  1749-4893},
  \urlprefix\url{http://www.nature.com/doifinder/10.1038/nphoton.2014.192}.

\bibitem[{\citenamefont{Tiecke et~al.}(2014)\citenamefont{Tiecke, Thompson,
  de~Leon, Liu, Vuletić, and Lukin}}]{tiecke_nanophotonic_2014}
\bibinfo{author}{\bibfnamefont{T.~G.} \bibnamefont{Tiecke}},
  \bibinfo{author}{\bibfnamefont{J.~D.} \bibnamefont{Thompson}},
  \bibinfo{author}{\bibfnamefont{N.~P.} \bibnamefont{de~Leon}},
  \bibinfo{author}{\bibfnamefont{L.~R.} \bibnamefont{Liu}},
  \bibinfo{author}{\bibfnamefont{V.}~\bibnamefont{Vuletić}}, \bibnamefont{and}
  \bibinfo{author}{\bibfnamefont{M.~D.} \bibnamefont{Lukin}},
  \bibinfo{journal}{Nature} \textbf{\bibinfo{volume}{508}},
  \bibinfo{pages}{241} (\bibinfo{year}{2014}), ISSN \bibinfo{issn}{0028-0836,
  1476-4687},
  \urlprefix\url{http://www.nature.com/doifinder/10.1038/nature13188}.

\bibitem[{\citenamefont{Lukin et~al.}(2001)\citenamefont{Lukin, Fleischhauer,
  Cote, Duan, Jaksch, Cirac, and Zoller}}]{lukin_dipole_2001}
\bibinfo{author}{\bibfnamefont{M.~D.} \bibnamefont{Lukin}},
  \bibinfo{author}{\bibfnamefont{M.}~\bibnamefont{Fleischhauer}},
  \bibinfo{author}{\bibfnamefont{R.}~\bibnamefont{Cote}},
  \bibinfo{author}{\bibfnamefont{L.~M.} \bibnamefont{Duan}},
  \bibinfo{author}{\bibfnamefont{D.}~\bibnamefont{Jaksch}},
  \bibinfo{author}{\bibfnamefont{J.~I.} \bibnamefont{Cirac}}, \bibnamefont{and}
  \bibinfo{author}{\bibfnamefont{P.}~\bibnamefont{Zoller}},
  \bibinfo{journal}{Physical Review Letters} \textbf{\bibinfo{volume}{87}},
  \bibinfo{pages}{037901} (\bibinfo{year}{2001}).

\bibitem[{\citenamefont{Saffman and Walker}(2002)}]{saffman_creating_2002}
\bibinfo{author}{\bibfnamefont{M.}~\bibnamefont{Saffman}} \bibnamefont{and}
  \bibinfo{author}{\bibfnamefont{T.~G.} \bibnamefont{Walker}},
  \bibinfo{journal}{Physical Review A} \textbf{\bibinfo{volume}{66}},
  \bibinfo{pages}{065403} (\bibinfo{year}{2002}).

\bibitem[{\citenamefont{Ningyuan et~al.}(2016)\citenamefont{Ningyuan,
  Georgakopoulos, Ryou, Schine, Sommer, and Simon}}]{ningyuan_observation_2016}
\bibinfo{author}{\bibfnamefont{J.}~\bibnamefont{Ningyuan}},
  \bibinfo{author}{\bibfnamefont{A.}~\bibnamefont{Georgakopoulos}},
  \bibinfo{author}{\bibfnamefont{A.}~\bibnamefont{Ryou}},
  \bibinfo{author}{\bibfnamefont{N.}~\bibnamefont{Schine}},
  \bibinfo{author}{\bibfnamefont{A.}~\bibnamefont{Sommer}}, \bibnamefont{and}
  \bibinfo{author}{\bibfnamefont{J.}~\bibnamefont{Simon}},
  \bibinfo{journal}{Physical Review A} \textbf{\bibinfo{volume}{93}},
  \bibinfo{pages}{041802} (\bibinfo{year}{2016}).

\bibitem[{\citenamefont{Ding and Shi}(2016)}]{ding_entanglement_2016}
\bibinfo{author}{\bibfnamefont{D.-S.} \bibnamefont{Ding}} \bibnamefont{and}
  \bibinfo{author}{\bibfnamefont{B.-S.} \bibnamefont{Shi}}, in
  \emph{\bibinfo{booktitle}{Lasers and {Electro}-{Optics} ({CLEO}), 2016
  {Conference} on}} (\bibinfo{publisher}{IEEE}, \bibinfo{year}{2016}), pp.
  \bibinfo{pages}{1--2}.

\bibitem[{\citenamefont{Pedersen and Mølmer}(2009)}]{pedersen_few_2009}
\bibinfo{author}{\bibfnamefont{L.~H.} \bibnamefont{Pedersen}} \bibnamefont{and}
  \bibinfo{author}{\bibfnamefont{K.}~\bibnamefont{Mølmer}},
  \bibinfo{journal}{Physical Review A} \textbf{\bibinfo{volume}{79}},
  \bibinfo{pages}{012320} (\bibinfo{year}{2009}).

\bibitem[{\citenamefont{Grankin et~al.}(2015)\citenamefont{Grankin, Brion,
  Bimbard, Boddeda, Usmani, Ourjoumtsev, and
  Grangier}}]{grankin_quantum-optical_2015}
\bibinfo{author}{\bibfnamefont{A.}~\bibnamefont{Grankin}},
  \bibinfo{author}{\bibfnamefont{E.}~\bibnamefont{Brion}},
  \bibinfo{author}{\bibfnamefont{E.}~\bibnamefont{Bimbard}},
  \bibinfo{author}{\bibfnamefont{R.}~\bibnamefont{Boddeda}},
  \bibinfo{author}{\bibfnamefont{I.}~\bibnamefont{Usmani}},
  \bibinfo{author}{\bibfnamefont{A.}~\bibnamefont{Ourjoumtsev}},
  \bibnamefont{and} \bibinfo{author}{\bibfnamefont{P.}~\bibnamefont{Grangier}},
  \bibinfo{journal}{Physical Review A} \textbf{\bibinfo{volume}{92}}
  (\bibinfo{year}{2015}), ISSN \bibinfo{issn}{1050-2947, 1094-1622},
  \urlprefix\url{https://link.aps.org/doi/10.1103/PhysRevA.92.043841}.

\bibitem[{\citenamefont{Wade et~al.}(2016)\citenamefont{Wade, Mattioli, and
  Mølmer}}]{wade_single-atom_2016}
\bibinfo{author}{\bibfnamefont{A.~C.~J.} \bibnamefont{Wade}},
  \bibinfo{author}{\bibfnamefont{M.}~\bibnamefont{Mattioli}}, \bibnamefont{and}
  \bibinfo{author}{\bibfnamefont{K.}~\bibnamefont{Mølmer}},
  \bibinfo{journal}{Physical Review A} \textbf{\bibinfo{volume}{94}},
  \bibinfo{pages}{053830} (\bibinfo{year}{2016}).

\bibitem[{\citenamefont{Maller et~al.}(2015)\citenamefont{Maller, Lichtman,
  Xia, Sun, Piotrowicz, Carr, Isenhower, and
  Saffman}}]{maller_rydberg-blockade_2015}
\bibinfo{author}{\bibfnamefont{K.~M.} \bibnamefont{Maller}},
  \bibinfo{author}{\bibfnamefont{M.~T.} \bibnamefont{Lichtman}},
  \bibinfo{author}{\bibfnamefont{T.}~\bibnamefont{Xia}},
  \bibinfo{author}{\bibfnamefont{Y.}~\bibnamefont{Sun}},
  \bibinfo{author}{\bibfnamefont{M.~J.} \bibnamefont{Piotrowicz}},
  \bibinfo{author}{\bibfnamefont{A.~W.} \bibnamefont{Carr}},
  \bibinfo{author}{\bibfnamefont{L.}~\bibnamefont{Isenhower}},
  \bibnamefont{and} \bibinfo{author}{\bibfnamefont{M.}~\bibnamefont{Saffman}},
  \bibinfo{journal}{Physical Review A} \textbf{\bibinfo{volume}{92}}
  (\bibinfo{year}{2015}), ISSN \bibinfo{issn}{1050-2947, 1094-1622},
  \urlprefix\url{https://link.aps.org/doi/10.1103/PhysRevA.92.022336}.

\bibitem[{\citenamefont{Jau et~al.}(2016)\citenamefont{Jau, Hankin, Keating,
  Deutsch, and Biedermann}}]{jau_entangling_2016}
\bibinfo{author}{\bibfnamefont{Y.-Y.} \bibnamefont{Jau}},
  \bibinfo{author}{\bibfnamefont{A.~M.} \bibnamefont{Hankin}},
  \bibinfo{author}{\bibfnamefont{T.}~\bibnamefont{Keating}},
  \bibinfo{author}{\bibfnamefont{I.~H.} \bibnamefont{Deutsch}},
  \bibnamefont{and} \bibinfo{author}{\bibfnamefont{G.~W.}
  \bibnamefont{Biedermann}}, \bibinfo{journal}{Nature Physics}
  \textbf{\bibinfo{volume}{12}}, \bibinfo{pages}{71} (\bibinfo{year}{2016}),
  ISSN \bibinfo{issn}{1745-2473, 1745-2481},
  \urlprefix\url{http://www.nature.com/articles/nphys3487}.

\bibitem[{\citenamefont{Firstenberg et~al.}(2013)\citenamefont{Firstenberg,
  Peyronel, Liang, Gorshkov, Lukin, and
  Vuletić}}]{firstenberg_attractive_2013}
\bibinfo{author}{\bibfnamefont{O.}~\bibnamefont{Firstenberg}},
  \bibinfo{author}{\bibfnamefont{T.}~\bibnamefont{Peyronel}},
  \bibinfo{author}{\bibfnamefont{Q.-Y.} \bibnamefont{Liang}},
  \bibinfo{author}{\bibfnamefont{A.~V.} \bibnamefont{Gorshkov}},
  \bibinfo{author}{\bibfnamefont{M.~D.} \bibnamefont{Lukin}}, \bibnamefont{and}
  \bibinfo{author}{\bibfnamefont{V.}~\bibnamefont{Vuletić}},
  \bibinfo{journal}{Nature} \textbf{\bibinfo{volume}{502}}, \bibinfo{pages}{71}
  (\bibinfo{year}{2013}), ISSN \bibinfo{issn}{0028-0836, 1476-4687},
  \urlprefix\url{http://www.nature.com/doifinder/10.1038/nature12512}.

\bibitem[{\citenamefont{Gorshkov et~al.}(2013)\citenamefont{Gorshkov, Nath, and
  Pohl}}]{gorshkov_dissipative_2013}
\bibinfo{author}{\bibfnamefont{A.~V.} \bibnamefont{Gorshkov}},
  \bibinfo{author}{\bibfnamefont{R.}~\bibnamefont{Nath}}, \bibnamefont{and}
  \bibinfo{author}{\bibfnamefont{T.}~\bibnamefont{Pohl}},
  \bibinfo{journal}{Physical Review Letters} \textbf{\bibinfo{volume}{110}}
  (\bibinfo{year}{2013}), ISSN \bibinfo{issn}{0031-9007, 1079-7114},
  \urlprefix\url{https://link.aps.org/doi/10.1103/PhysRevLett.110.153601}.

\bibitem[{\citenamefont{Gardiner}(1993)}]{gardiner_driving_1993}
\bibinfo{author}{\bibfnamefont{C.~W.} \bibnamefont{Gardiner}},
  \bibinfo{journal}{Phys. Rev. Lett.} \textbf{\bibinfo{volume}{70}},
  \bibinfo{pages}{2269} (\bibinfo{year}{1993}).

\bibitem[{\citenamefont{Pedersen et~al.}(2007)\citenamefont{Pedersen, Møller,
  and Mølmer}}]{pedersen_fidelity_2007}
\bibinfo{author}{\bibfnamefont{L.~H.} \bibnamefont{Pedersen}},
  \bibinfo{author}{\bibfnamefont{N.~M.} \bibnamefont{Møller}},
  \bibnamefont{and} \bibinfo{author}{\bibfnamefont{K.}~\bibnamefont{Mølmer}},
  \bibinfo{journal}{Phys. Lett. A} \textbf{\bibinfo{volume}{367}},
  \bibinfo{pages}{47} (\bibinfo{year}{2007}).

\bibitem[{\citenamefont{Brennecke et~al.}(2007)\citenamefont{Brennecke, Donner,
  Ritter, Bourdel, Köhl, and Esslinger}}]{brennecke_cavity_2007}
\bibinfo{author}{\bibfnamefont{F.}~\bibnamefont{Brennecke}},
  \bibinfo{author}{\bibfnamefont{T.}~\bibnamefont{Donner}},
  \bibinfo{author}{\bibfnamefont{S.}~\bibnamefont{Ritter}},
  \bibinfo{author}{\bibfnamefont{T.}~\bibnamefont{Bourdel}},
  \bibinfo{author}{\bibfnamefont{M.}~\bibnamefont{Köhl}}, \bibnamefont{and}
  \bibinfo{author}{\bibfnamefont{T.}~\bibnamefont{Esslinger}},
  \bibinfo{journal}{Nature} \textbf{\bibinfo{volume}{450}},
  \bibinfo{pages}{268} (\bibinfo{year}{2007}), ISSN \bibinfo{issn}{0028-0836,
  1476-4687},
  \urlprefix\url{http://www.nature.com/doifinder/10.1038/nature06120}.

\bibitem[{\citenamefont{Murch et~al.}(2008)\citenamefont{Murch, Moore, Gupta,
  and Stamper-Kurn}}]{murch_observation_2008}
\bibinfo{author}{\bibfnamefont{K.~W.} \bibnamefont{Murch}},
  \bibinfo{author}{\bibfnamefont{K.~L.} \bibnamefont{Moore}},
  \bibinfo{author}{\bibfnamefont{S.}~\bibnamefont{Gupta}}, \bibnamefont{and}
  \bibinfo{author}{\bibfnamefont{D.~M.} \bibnamefont{Stamper-Kurn}},
  \bibinfo{journal}{Nature Physics} \textbf{\bibinfo{volume}{4}},
  \bibinfo{pages}{561} (\bibinfo{year}{2008}), ISSN \bibinfo{issn}{1745-2473,
  1745-2481}, \urlprefix\url{http://www.nature.com/articles/nphys965}.

\bibitem[{\citenamefont{Greene et~al.}(2000)\citenamefont{Greene, Dickinson,
  and Sadeghpour}}]{greene_creation_2000}
\bibinfo{author}{\bibfnamefont{C.~H.} \bibnamefont{Greene}},
  \bibinfo{author}{\bibfnamefont{A.~S.} \bibnamefont{Dickinson}},
  \bibnamefont{and} \bibinfo{author}{\bibfnamefont{H.~R.}
  \bibnamefont{Sadeghpour}}, \bibinfo{journal}{Physical Review Letters}
  \textbf{\bibinfo{volume}{85}}, \bibinfo{pages}{2458} (\bibinfo{year}{2000}),
  ISSN \bibinfo{issn}{0031-9007, 1079-7114},
  \urlprefix\url{https://link.aps.org/doi/10.1103/PhysRevLett.85.2458}.

\bibitem[{\citenamefont{Boisseau et~al.}(2002)\citenamefont{Boisseau, Simbotin,
  and Côté}}]{boisseau_macrodimers:_2002}
\bibinfo{author}{\bibfnamefont{C.}~\bibnamefont{Boisseau}},
  \bibinfo{author}{\bibfnamefont{I.}~\bibnamefont{Simbotin}}, \bibnamefont{and}
  \bibinfo{author}{\bibfnamefont{R.}~\bibnamefont{Côté}},
  \bibinfo{journal}{Physical Review Letters} \textbf{\bibinfo{volume}{88}}
  (\bibinfo{year}{2002}), ISSN \bibinfo{issn}{0031-9007, 1079-7114},
  \urlprefix\url{https://link.aps.org/doi/10.1103/PhysRevLett.88.133004}.

\bibitem[{\citenamefont{Khani et~al.}(2009)\citenamefont{Khani, Gambetta,
  Motzoi, and Wilhelm}}]{khani_optimal_2009}
\bibinfo{author}{\bibfnamefont{B.}~\bibnamefont{Khani}},
  \bibinfo{author}{\bibfnamefont{J.~M.} \bibnamefont{Gambetta}},
  \bibinfo{author}{\bibfnamefont{F.}~\bibnamefont{Motzoi}}, \bibnamefont{and}
  \bibinfo{author}{\bibfnamefont{F.~K.} \bibnamefont{Wilhelm}},
  \bibinfo{journal}{Physica Scripta} \textbf{\bibinfo{volume}{T137}},
  \bibinfo{pages}{014021 (5pp)} (\bibinfo{year}{2009}),
  \urlprefix\url{http://stacks.iop.org/1402-4896/T137/014021}.

\bibitem[{\citenamefont{Khani et~al.}(2012)\citenamefont{Khani, Merkel, Motzoi,
  Gambetta, and Wilhelm}}]{khani_high-fidelity_2012}
\bibinfo{author}{\bibfnamefont{B.}~\bibnamefont{Khani}},
  \bibinfo{author}{\bibfnamefont{S.~T.} \bibnamefont{Merkel}},
  \bibinfo{author}{\bibfnamefont{F.}~\bibnamefont{Motzoi}},
  \bibinfo{author}{\bibfnamefont{J.~M.} \bibnamefont{Gambetta}},
  \bibnamefont{and} \bibinfo{author}{\bibfnamefont{F.~K.}
  \bibnamefont{Wilhelm}}, \bibinfo{journal}{Physical Review A}
  \textbf{\bibinfo{volume}{85}} (\bibinfo{year}{2012}), ISSN
  \bibinfo{issn}{1050-2947, 1094-1622},
  \urlprefix\url{https://link.aps.org/doi/10.1103/PhysRevA.85.022306}.

\bibitem[{\citenamefont{Abdumalikov et~al.}(2010)\citenamefont{Abdumalikov,
  Astafiev, Zagoskin, Pashkin, Nakamura, and
  Tsai}}]{abdumalikov_electromagnetically_2010}
\bibinfo{author}{\bibfnamefont{A.~A.} \bibnamefont{Abdumalikov}},
  \bibinfo{author}{\bibfnamefont{O.}~\bibnamefont{Astafiev}},
  \bibinfo{author}{\bibfnamefont{A.~M.} \bibnamefont{Zagoskin}},
  \bibinfo{author}{\bibfnamefont{Y.~A.} \bibnamefont{Pashkin}},
  \bibinfo{author}{\bibfnamefont{Y.}~\bibnamefont{Nakamura}}, \bibnamefont{and}
  \bibinfo{author}{\bibfnamefont{J.~S.} \bibnamefont{Tsai}},
  \bibinfo{journal}{Physical Review Letters} \textbf{\bibinfo{volume}{104}}
  (\bibinfo{year}{2010}), ISSN \bibinfo{issn}{0031-9007, 1079-7114},
  \urlprefix\url{https://link.aps.org/doi/10.1103/PhysRevLett.104.193601}.

\bibitem[{\citenamefont{Murali et~al.}(2004)\citenamefont{Murali, Dutton,
  Oliver, Crankshaw, and Orlando}}]{murali_probing_2004}
\bibinfo{author}{\bibfnamefont{K.}~\bibnamefont{Murali}},
  \bibinfo{author}{\bibfnamefont{Z.}~\bibnamefont{Dutton}},
  \bibinfo{author}{\bibfnamefont{W.}~\bibnamefont{Oliver}},
  \bibinfo{author}{\bibfnamefont{D.}~\bibnamefont{Crankshaw}},
  \bibnamefont{and} \bibinfo{author}{\bibfnamefont{T.}~\bibnamefont{Orlando}},
  \bibinfo{journal}{Physical Review Letters} \textbf{\bibinfo{volume}{93}}
  (\bibinfo{year}{2004}), ISSN \bibinfo{issn}{0031-9007, 1079-7114},
  \urlprefix\url{https://link.aps.org/doi/10.1103/PhysRevLett.93.087003}.

\bibitem[{\citenamefont{Theis et~al.}(2016)\citenamefont{Theis, Motzoi, and
  Wilhelm}}]{theis_simultaneous_2016}
\bibinfo{author}{\bibfnamefont{L.~S.} \bibnamefont{Theis}},
  \bibinfo{author}{\bibfnamefont{F.}~\bibnamefont{Motzoi}}, \bibnamefont{and}
  \bibinfo{author}{\bibfnamefont{F.~K.} \bibnamefont{Wilhelm}},
  \bibinfo{journal}{Phys. Rev. A} \textbf{\bibinfo{volume}{93}},
  \bibinfo{pages}{012324} (\bibinfo{year}{2016}).

\bibitem[{\citenamefont{Goerz et~al.}(2017)\citenamefont{Goerz, Motzoi, Whaley,
  and Koch}}]{goerz_charting_2017}
\bibinfo{author}{\bibfnamefont{M.~H.} \bibnamefont{Goerz}},
  \bibinfo{author}{\bibfnamefont{F.}~\bibnamefont{Motzoi}},
  \bibinfo{author}{\bibfnamefont{K.~B.} \bibnamefont{Whaley}},
  \bibnamefont{and} \bibinfo{author}{\bibfnamefont{C.~P.} \bibnamefont{Koch}},
  \bibinfo{journal}{npj Quantum Information} \textbf{\bibinfo{volume}{3}},
  \bibinfo{pages}{37} (\bibinfo{year}{2017}), ISSN \bibinfo{issn}{2056-6387},
  \urlprefix\url{https://doi.org/10.1038/s41534-017-0036-0}.

\end{thebibliography}
\end{document}